\documentclass{aa}

\usepackage{linenoaa}
\usepackage{hyperref}
\usepackage{siunitx}
\usepackage{booktabs}
\usepackage{tabularx}
\usepackage{array}
\usepackage{graphicx}
\usepackage{subcaption}
\usepackage{afterpage}
\usepackage{amsmath}
\usepackage{upgreek}
\usepackage{mathrsfs}
\usepackage{CJKutf8}
\usepackage{ulem}
\usepackage{multirow}
\usepackage{booktabs}

\usepackage{txfonts}
\usepackage{hyperref}

\newcommand{\deriv}[2]{\dfrac{{\rm d} #1}{{\rm d} #2}}
\newcommand{\aderiv}[2]{\dfrac{{\Delta} #1}{{\Delta} #2}}

\begin{document} 

\title{
Dynamically New Comet C/2025 D1 (Groeller) with Record Perihelion Distance
}

\author{
Man-To Hui 
\begin{CJK}{UTF8}{bsmi}
(許文韜)
\end{CJK}\inst{1}
\and
Robert Weryk\inst{2}
\and
Marco Micheli\inst{3}
\and
Sam Deen\inst{4}
\and
David J. Tholen\inst{5}
\and
Jianchun Shi 
\begin{CJK}{UTF8}{bsmi}
(史建春)
\end{CJK}\inst{1}
\and
\begin{CJK}{UTF8}{bsmi}
Xian Shi 
(史弦)
\end{CJK}\inst{1}
\and
Richard Wainscoat\inst{5}
}
\institute{
Shanghai Astronomical Observatory, Chinese Academy of Sciences,
No. 80 Rd Nandan, Shanghai 200030, Mainland China
\and
Physics and Astronomy, The University of Western Ontario,
1151 Richmond Street,
London ON N6A 3K7, Canada
\and
ESA NEO Coordination Centre, Planetary Defence Office, Largo Galileo Galilei 1, I-00044 Frascati (RM), Italy
\and
Simi Valley, CA, USA
\and
Institute for Astronomy, University of Hawai`i, 
2680 Woodlawn Drive, Honolulu, HI 96822, USA
\\
\email{
{mthui@shao.ac.cn,
manto@hawaii.edu}
}}

\date{}


\abstract
{
We studied C/2025 D1 (Groeller), a long-period comet with an unprecedented perihelion distance of 14.1 au, using archival observations. The data reveals that it had been active at inbound heliocentric distances $r_{\rm H} \gtrsim 20$ au. Initially, the comet intrinsically brightened at $r_{\rm H} \gtrsim 16$ au, with brightening parameters comparable to those of other long-period comets. However, observations after late 2023 showed a gradual decay, despite the inbound trajectory of the comet. To our knowledge, such behaviours have not been observed for other long-period comets at similar heliocentric distances. We speculate that this might be linked to the onset of CO$_{2}$ sublimation and/or crystallisation processes. Alternatively, the activity source might have been exhausted. The surface brightness profile of the coma indicates a steady-state mass loss, implying supervolatile sublimation as the primary driver of the observed activity. Despite changes in the orbital plane angle, the circularly symmetric coma persisted throughout the observed period, indicative of the dominance of large grains in the coma. Assuming the activity trend is independent of bandpass, we found that comet was redder than many other solar system comets. Our model-dependent constraint estimates the nucleus radius to be $\gtrsim\!0.4$ km. We performed astrometric measurements, refined the orbital solution, and derived the original and future orbits of the comet. Our N-body integration, accounting for the Galactic tide, strongly favours that the comet is dynamically new, with its previous perihelion at $\gtrsim\!60$ au from the Sun $\gtrsim\!6$ Myr ago. It is highly likely that the comet will be lost from our solar system after the current apparition.
}


\authorrunning{M.-T. Hui et al.}
\maketitle
%

\nolinenumbers

\section{Introduction}
\label{sec_intro}

Comets are volatile-rich relics formed and survived from the violent and chaotic stages of the early solar system approximately 4.5 Gyr ago. The present-day solar system hosts two primary cometary reservoirs -- the Kuiper Belt and the Oort Cloud, which supply short- and long-period comets, respectively. In general, short-period comets have heliocentric orbits with modest eccentricities and inclinations with respect to the ecliptic, whereas long-period comets travel on nearly parabolic trajectories from isotropic directions.

Compared to short-period comets, the long-period counterparts may be even more pristine, as they have resided predominantly in the deep-freeze environment with an ambient temperature of $\sim\!10$ K since their implantation in the Oort Cloud, despite that both types of comets were mostly formed in a massive primordial trans-Neptunian disc, with a small fraction born closer to the Sun \citep[e.g.,][]{2017ApJ...845...27N,2019AJ....157..181V}. Long-period comets can be classified as dynamically new or old, depending on whether they have previously entered the planetary region \citep[$\la\!15$ au;][]{2010MNRAS.404.1886K}. Although long-period comets, particularly dynamically new ones, are considered among the most pristine small solar system bodies, they are not immune to thermal processes, as evidenced by their mass-loss activity. While most comets exhibit activity due to water-ice sublimation within heliocentric distances $\la\!5$ au \citep[e.g.,][]{1950ApJ...111..375W}, others can be active at $10 \la r_{\rm H} \la 16$ au attributed to crystallisation of amorphous ice \citep{2012AJ....144...97G} or at even greater heliocentric distances because of supervolatile (e.g., CO, CO$_{2}$) sublimation \citep[e.g.,][]{2021AJ....161..188J}. Even objects as far out as in the Oort Cloud may undergo substantial thermophysical processing due to cosmic ray bombardment \citep{2020ApJ...890...89G, 2020ApJ...901..136M}, which may trigger cometary outbursts therein \citep{2024Icar..41516066B}. Despite these processes, long-period comets, especially dynamically new ones with large perihelion distances, remain scientifically significant for understanding the evolutionary pathways of the solar system.

Comet C/2025 D1 (Groeller) was discovered on 2025 February 20 at an inbound heliocentric distance of $r_{\rm H} = 15.1$ au. Serendipitous prediscovery observations dating back to as early as 2018 were identified, facilitating robust orbit determination \citep{2025MPEC....D...83W}. According to JPL Horizons' orbital solution, C/2025 D1 has a perihelion distance of $q = 14.1$ au, the largest known among comets. With a slightly hyperbolic heliocentric eccentricity ($e = 1.003$) on the inbound leg of the trajectory due to perturbations of the giant planets, and an orbital inclination nearly orthogonal to the ecliptic ($i = 84\fdg5$), the perihelion passage of the comet will occur in 2028 May. In this paper, we analyse archival observations to examine comet C/2025 D1 on the inbound leg of its trajectory.

\section{Observations}
\label{sec_obs}

\begin{figure*}
\centering
\includegraphics[width=1\linewidth]{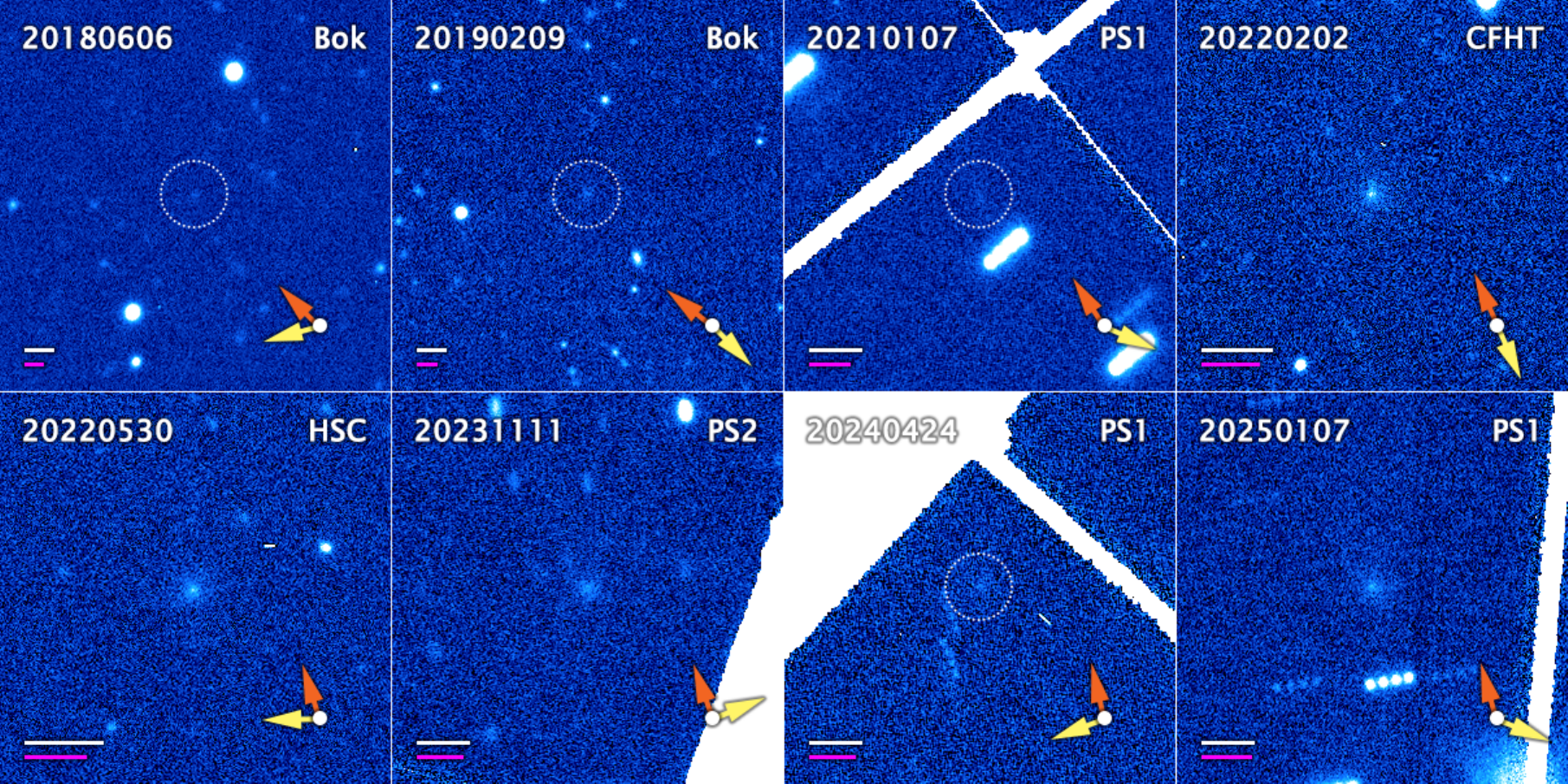}
\caption{Selected archival observations of comet C/2025 D1 (Groeller). In some of the panels where the comet appears faint, a white dotted circle is used to mark its location. The white and magenta scale bars represent an angular distance of 10\arcsec~and a linear distance of $10^5$ km projected at the observer-centric distance of the comet, respectively. J2000 equatorial north is up and east is to the left. In each panel, the antisolar direction (yellow arrow) and the projected negative heliocentric velocity of the comet in the observer's sky plane (dark orange arrow) are also indicated.}
\label{fig:obs}
\end{figure*}

We used the Solar System Object Search tool \citep{2012PASP..124..579G} at the Canadian Astronomy Data Centre (CADC)\footnote{\url{https://www.cadc-ccda.hia-iha.nrc-cnrc.gc.ca/en/ssois/}} to obtain archival observations of comet C/2025 D1. After visually inspecting the downloaded data, we robustly identified the comet in images from three facilities: the 2.3 m Bok telescope with a four $4032 \times 4096$ chip CCD at the Kitt Peak National Observatory, the 3.6 m Canada-France-Hawaii Telescope (CFHT) with MegaCam \citep{2003SPIE.4841..513A}, and the 8.4 m Subaru telescope with Hyper Suprime-Cam \citep[HSC;][]{2010AIPC.1279..120T}, both located at the summit of Mauna Kea, Hawai`i. The image pixel scales of the these facilities are 0\farcs453, 0\farcs187, and 0\farcs166, respectively, all in unbinned mode. In addition to the CADC query, we searched for data taken from the two 1.8 m Pan-STARRS survey telescopes \citep[PS1 and PS2; ][]{2016arXiv161205560C} at Haleakal{\=a}, Hawai`i, with an image scale of 0\farcs25 pixel$^{-1}$, and successfully identified the comet therein. The observing information and the viewing geometry of the comet are summarised in Table \ref{tab:vgeo}, and Figure \ref{fig:obs} displays selected images of the comet from the archival data.

Except for the CFHT data, the Word Coordinate System (WCS) header information of the archival data was created or updated using {\tt astrometry.net} \citep{2010AJ....139.1782L} in combination with the Gaia Data Release 2 \citep[DR2;][]{2018A&A...616A...1G} and Tycho-2 \citep{2000A&A...355L..27H} catalogues, as catalogued sources were noticed to be visually offset from the observed field sources in {\tt SAOImage DS9} \citep{2003ASPC..295..489J}. The WCS header information updated by {\tt astrometry.net} served as a preliminary astrometric calibration for subsequent procedures. All images queried from the CADC, except for a single HSC image (see Table \ref{tab:vgeo}), were further astrometrically calibrated with the Gaia DR3 \citep{2023A&A...674A...1G} in {\tt Astrometrica} \citep{2012ascl.soft03012R}, yielding the astrometry and associated measurement uncertainties for the comet. The HSC image, affected by significant vignetting from an edge shadow even after standard calibration, exhibited substantial field distortion as a result of diffraction, which baffled {\tt Astrometrica}. Therefore, we followed the detailed methodology described in \citet{2023PSJ.....4..215H} in combination with {\tt AstroMagic}\footnote{\url{http://www.astromagic.it/eng/astromagic.html}} to visualise pattern matching while account for the field distortion coefficients specific to the HSC CCD chip. This approach enabled successful measurement of the comet's astrometry and associated error in this HSC image. Using the same technique applied in \citet{2024AJ....167..140H} and \citet{2024ApJ...975L...3H}, we measured the astrometry of the comet in the PS images, which were already astrometrically and photometrically calibrated \citep{2020ApJS..251....4W}.

Although the non-PS images were photometrically calibrated by {\tt Astrometrica} during the astrometric reduction, the results were preliminary, necessitating refined photometric calibration. To improve the calibration, we measured fluxes of field stars using a circular aperture with a radius equal to twice the mean seeing FWHM value of field stars and estimated the sky background from a concentric annulus with inner and outer radii of twice and four times the photometric aperture radius, respectively. Referencing ATLAS Refcat2 \citep{2018ApJ...867..105T}, we determined the zero-point (and the colour term, if statistically significant) for each image, best fitted using {\tt MPFIT} \citep{2009ASPC..411..251M}, as done in our previous work \citep[e.g.,][]{2024AJ....167..140H}. To minimise potential effects from the varying observing geometry of C/2025 D1, we performed photometry of the comet using circular apertures with fixed linear radii projected at its observer-centric distance, ranging from $2.5 \times 10^4$ to $4 \times 10^4$ km in increments of 5000 km. The sky background was estimated from a concentric annulus with inner and outer radii equal to twice and four times, respectively, the angular radius of the largest circular aperture used for comet photometry. Photometry of the comet in the PS data was more straightforward, as there was no need to estimate the sky background, and we adopted the same fixed apertures using the method described in \citet{2024AJ....167..140H}. For non-PS images, photometric uncertainties were propagated from errors in the photometric recalibration and flux measurement uncertainties, assuming Poisson statistics. However, for the PS data, which were sky-subtracted, Poisson statistics were inapplicable. Fortunately, most PS-observed epochs had multiple measurements in the same filter from the same night (see Table \ref{tab:vgeo}), allowing us to use the standard deviation of repeated measurements as the measurement error. The few PS measurement singletons were discarded.

\longtab[1]{
\begin{longtable}{cccccccccccc}
\caption{
Archival Observations and Viewing Geometry of Comet C/2025 D1 (Groeller)
\label{tab:vgeo}
}
\\
\toprule
Date & \multicolumn{4}{c}{Archival Observations} &
\multicolumn{7}{c}{Viewing Geometry} \\ \cmidrule(lr){2-5} \cmidrule(lr){6-12}
(UTC) & Facility & Filter & \# images & Exposure (s) & $r_{\rm H}$ (au)$^{a}$ & $\Delta$ (au)$^{b}$ & $\alpha$ (\degr)$^{c}$ & $\varepsilon$ (\degr)$^{d}$ & $\theta_{-\odot}$ (\degr)$^{e}$ & $\theta_{-{\bf V}}$ (\degr)$^{f}$ & $\psi$ (\degr)$^{g}$
\\
\midrule
\endfirsthead
\caption{Continued.} \\
\toprule
Date & \multicolumn{4}{c}{Archival Observations} &
\multicolumn{7}{c}{Viewing Geometry} 
\\ \cmidrule(lr){2-5} \cmidrule(lr){6-12}
(UTC) & Facility & Filter & \# images & Exposure (s) & 
$r_{\rm H}$ (au)$^{a}$ & $\Delta$ (au)$^{b}$ & $\alpha$ (\degr)$^{c}$ & $\varepsilon$ (\degr)$^{d}$ & $\theta_{-\odot}$ (\degr)$^{e}$ & $\theta_{-{\bf V}}$ (\degr)$^{f}$ & $\psi$ (\degr)$^{g}$
\\
\midrule
\endhead
\hline
\endfoot
\hline
\endlastfoot
2018-06-06 & Bok & $g$ & 1 & 250 & 21.313 & 21.637 & 2.6 & 70.1 & 105.1 & 44.8 & +2.3 \\
2018-06-07 & Bok & $g$ & 1$^{\dagger}$ & 197 & 21.309 & 21.639 & 2.6 & 69.8 & 104.1 & 44.7 & +2.2 \\
2018-06-09 & Bok & $g$ & 1 & 250 & 21.303 & 21.643 & 2.6 & 69.2 & 102.1 & 44.6 & +2.2 \\
2018-06-18 & Bok & $r$ & 1 & 210 & 21.276 & 21.658 & 2.5 & 66.7 & 93.0 & 44.0 & +1.9 \\
2019-01-28 & Bok & $r$ & 1 & 152 & 20.594 & 20.148 & 2.5 & 115.6 & 239.0 & 53.3 & -0.2 \\
2019-02-09 & Bok & $r$ & 1 & 156 & 20.557 & 20.129 & 2.5 & 114.5 & 223.3 & 51.7 & +0.4 \\
2019-02-12 & Bok & $r$ & 1 & 181 & 20.549 & 20.128 & 2.5 & 114.0 & 219.7 & 51.3 & +0.5 \\
2020-12-06 & PS1 & $i$ & 4 &  45 & 18.623 & 18.269 & 2.9 & 109.6 & 281.8 & 33.9 & -2.6 \\
2021-01-02 & PS1 & $i$ & 4 &  45 & 18.548 & 18.014 & 2.6 & 121.5 & 252.1 & 33.2 & -1.6 \\
2021-01-04 & PS1 & $i$ & 4 &  45 & 18.542 & 18.000 & 2.6 & 122.2 & 249.6 & 33.0 & -1.5 \\
2021-01-06 & PS1 & $i$ & 4 &  45 & 18.537 & 17.987 & 2.6 & 122.7 & 247.1 & 32.9 & -1.4 \\
2021-01-07 & PS1 & $i$ & 4 &  45 & 18.534 & 17.980 & 2.6 & 123.0 & 245.7 & 32.9 & -1.3 \\
2021-02-23 & PS1 & $i$ & 6 &  45 & 18.405 & 17.883 & 2.7 & 120.5 & 181.2 & 28.1 & +1.2 \\
2021-02-24 & PS1 & $i$ & 4 &  45 & 18.402 & 17.886 & 2.7 & 120.1 & 179.9 & 28.0 & +1.2 \\
2021-03-21 & PS1 & $i$ & 3 &  45 & 18.334 & 17.995 & 3.0 & 108.4 & 151.3 & 24.9 & +2.3 \\
2021-03-23 & PS1 & $i$ & 4 &  45 & 18.329 & 18.007 & 3.0 & 107.4 & 149.4 & 24.7 & +2.4 \\
2021-03-24 & PS1 & $i$ & 2 &  45 & 18.326 & 18.013 & 3.0 & 106.8 & 148.3 & 24.6 & +2.4 \\
2021-03-25 & PS1 & $i$ & 4 &  45 & 18.323 & 18.020 & 3.0 & 106.3 & 147.3 & 24.5 & +2.5 \\
2021-03-26 & PS1 & $i$ & 4 &  45 & 18.320 & 18.026 & 3.0 & 105.6 & 146.3 & 24.4 & +2.5 \\
2021-04-02 & PS1 & $i$ & 4 &  45 & 18.301 & 18.076 & 3.1 & 101.4 & 139.4 & 23.7 & +2.7 \\
2021-04-20 & PS1 & $i$ & 2 &  45 & 18.253 & 18.213 & 3.2 & 90.7 & 123.9 & 22.2 & +3.1 \\
2021-11-21 & PS1 & $i$ & 4 &  45 & 17.680 & 17.452 & 3.1 & 101.8 & 290.4 & 27.4 & -3.1 \\
2021-11-24 & PS2 & $i$ & 4 &  45 & 17.672 & 17.411 & 3.1 & 103.8 & 288.0 & 27.4 & -3.1 \\
2021-11-26 & PS1 & $w$ & 2 &  45 & 17.667 & 17.384 & 3.1 & 105.1 & 286.4 & 27.4 & -3.0 \\
2022-02-02 & CFHT & $r$ & 1 & 100 & 17.492 & 16.825 & 2.4 & 131.4 & 203.7 & 24.2 & +0.0 \\
2022-05-30 & HSC & $g$ & 1$^{\ddagger}$ & 120 & 17.197 & 17.598 & 3.1 & 65.2 & 90.9 & 17.2 & +3.0 \\
 & & & 1 &  90 & 17.197 & 17.598 & 3.1 & 65.1 & 90.8 & 17.2 & +3.0 \\
2022-10-30 & PS2 & $g$ & 1 &  27 & 16.824 & 16.869 & 3.4 & 85.7 & 302.0 & 22.4 & -3.3 \\
2022-10-30 & PS2 & $r$ & 1 &  27 & 16.824 & 16.869 & 3.4 & 85.7 & 302.0 & 22.4 & -3.3 \\
2022-11-02 & PS2 & $g$ & 1 &  27 & 16.817 & 16.823 & 3.4 & 87.9 & 300.1 & 22.5 & -3.4 \\
2022-11-02 & PS2 & $i$ & 1 &  27 & 16.817 & 16.823 & 3.4 & 87.9 & 300.1 & 22.5 & -3.4 \\
2022-11-06 & PS2 & $i$ & 1 &  27 & 16.807 & 16.761 & 3.4 & 91.0 & 297.5 & 22.5 & -3.4 \\
2022-11-06 & PS2 & $r$ & 1 &  27 & 16.807 & 16.761 & 3.4 & 91.0 & 297.5 & 22.5 & -3.4 \\
2022-11-10 & PS2 & $i$ & 4 &  45 & 16.798 & 16.700 & 3.4 & 94.0 & 294.9 & 22.6 & -3.4 \\
2022-11-12 & PS1 & $i$ & 4 &  45 & 16.793 & 16.669 & 3.4 & 95.5 & 293.6 & 22.6 & -3.4 \\
2022-11-12 & PS2 & $i$ & 4 &  45 & 16.793 & 16.669 & 3.4 & 95.5 & 293.6 & 22.6 & -3.4 \\
2022-11-12 & PS2 & $r$ & 1 &  27 & 16.793 & 16.669 & 3.4 & 95.5 & 293.6 & 22.6 & -3.4 \\
2022-11-12 & PS2 & $y$ & 1 &  27 & 16.793 & 16.669 & 3.4 & 95.5 & 293.6 & 22.6 & -3.4 \\
2022-11-14 & PS2 & $i$ & 4 &  45 & 16.788 & 16.639 & 3.4 & 97.0 & 292.3 & 22.6 & -3.4 \\
2022-11-19 & PS2 & $g$ & 1 &  27 & 16.776 & 16.563 & 3.3 & 100.8 & 288.8 & 22.6 & -3.3 \\
2022-11-19 & PS2 & $i$ & 1 &  27 & 16.776 & 16.563 & 3.3 & 100.8 & 288.8 & 22.6 & -3.3 \\
2022-12-02 & PS2 & $g$ & 1 &  27 & 16.746 & 16.375 & 3.2 & 110.5 & 279.1 & 22.6 & -3.1 \\
2022-12-02 & PS2 & $y$ & 1 &  27 & 16.746 & 16.375 & 3.2 & 110.5 & 279.1 & 22.6 & -3.1 \\
2022-12-14 & PS1 & $i$ & 4 &  45 & 16.717 & 16.217 & 2.9 & 119.1 & 268.7 & 22.3 & -2.7 \\
2022-12-14 & PS2 & $i$ & 4 &  45 & 16.717 & 16.217 & 2.9 & 119.1 & 268.7 & 22.3 & -2.7 \\
2023-01-09 & PS1 & $i$ & 4 &  45 & 16.657 & 15.960 & 2.4 & 133.9 & 238.2 & 21.3 & -1.4 \\
2023-03-30 & PS1 & $i$ & 4 &  45 & 16.474 & 16.114 & 3.3 & 109.5 & 131.0 & 15.9 & +2.9 \\
2023-05-02 & PS1 & $i$ & 4 &  45 & 16.399 & 16.467 & 3.5 & 84.4 & 108.6 & 14.7 & +3.5 \\
2023-05-02 & PS2 & $i$ & 4 &  45 & 16.399 & 16.468 & 3.5 & 84.3 & 108.6 & 14.7 & +3.5 \\
2023-11-11 & PS2 & $w$ & 4 &  45 & 15.981 & 15.877 & 3.5 & 94.3 & 291.7 & 19.0 & -3.5 \\
2024-01-01 & PS1 & $i$ & 4 &  45 & 15.877 & 15.168 & 2.5 & 134.8 & 251.1 & 18.1 & -2.0 \\
2024-01-03 & PS2 & $w$ & 4 &  45 & 15.873 & 15.150 & 2.5 & 136.1 & 248.6 & 18.0 & -1.9 \\
2024-01-22 & PS1 & $i$ & 4 &  45 & 15.834 & 15.020 & 2.1 & 144.8 & 218.3 & 17.1 & -0.7 \\
2024-01-23 & PS2 & $i$ & 4 &  45 & 15.832 & 15.016 & 2.0 & 145.0 & 216.4 & 17.1 & -0.7 \\
2024-01-24 & PS2 & $i$ & 4 &  45 & 15.830 & 15.012 & 2.0 & 145.2 & 214.4 & 17.0 & -0.6 \\
2024-03-24 & PS1 & $i$ & 1 &  45 & 15.712 & 15.237 & 3.2 & 116.8 & 129.6 & 13.8 & +2.9 \\
2024-03-24 & PS2 & $i$ & 1 &  45 & 15.712 & 15.237 & 3.2 & 116.8 & 129.6 & 13.8 & +2.9 \\
2024-03-27 & PS2 & $i$ & 4 &  45 & 15.706 & 15.269 & 3.3 & 114.3 & 127.4 & 13.7 & +3.0 \\
2024-04-19 & PS2 & $i$ & 8 &  45 & 15.661 & 15.539 & 3.7 & 95.2 & 113.4 & 13.0 & +3.6 \\
2024-04-24 & PS1 & $i$ & 4 &  45 & 15.651 & 15.603 & 3.7 & 90.9 & 110.7 & 12.9 & +3.6 \\
2024-04-25 & PS2 & $i$ & 4 &  45 & 15.649 & 15.616 & 3.7 & 90.1 & 110.2 & 12.9 & +3.6 \\
2024-04-26 & PS2 & $i$ & 4 &  45 & 15.648 & 15.628 & 3.7 & 89.3 & 109.7 & 12.8 & +3.7 \\
2024-05-21 & PS2 & $i$ & 4 &  45 & 15.600 & 15.939 & 3.5 & 68.7 & 97.3 & 12.7 & +3.5 \\
2024-11-19 & PS2 & $i$ & 4 &  45 & 15.271 & 15.038 & 3.6 & 101.8 & 285.8 & 16.3 & -3.6 \\
2024-11-27 & PS2 & $z$ & 1 & 120 & 15.257 & 14.904 & 3.5 & 109.2 & 281.8 & 16.2 & -3.5 \\
2024-12-02 & PS1 & $w$ & 4 &  45 & 15.248 & 14.825 & 3.4 & 113.7 & 279.1 & 16.2 & -3.4 \\
2024-12-05 & PS1 & $z$ & 1 & 120 & 15.243 & 14.778 & 3.3 & 116.5 & 277.3 & 16.1 & -3.3 \\
2024-12-05 & PS2 & $z$ & 1 & 120 & 15.243 & 14.778 & 3.3 & 116.5 & 277.3 & 16.1 & -3.3 \\
2024-12-09 & PS1 & $y$ & 1 & 120 & 15.237 & 14.718 & 3.2 & 120.2 & 274.9 & 16.1 & -3.1 \\
2024-12-09 & PS1 & $z$ & 1 & 120 & 15.237 & 14.718 & 3.2 & 120.2 & 274.9 & 16.1 & -3.1 \\
2024-12-14 & PS1 & $i$ & 4 &  45 & 15.228 & 14.648 & 3.1 & 124.6 & 271.5 & 16.0 & -2.9 \\
2024-12-15 & PS1 & $i$ & 4 &  45 & 15.226 & 14.634 & 3.0 & 125.5 & 270.8 & 15.9 & -2.9 \\
2024-12-15 & PS2 & $i$ & 4 &  45 & 15.227 & 14.635 & 3.0 & 125.5 & 270.8 & 15.9 & -2.9 \\
2024-12-17 & PS2 & $i$ & 4 &  45 & 15.223 & 14.608 & 3.0 & 127.2 & 269.4 & 15.9 & -2.8 \\
2024-12-22 & PS1 & $w$ & 8 &  45 & 15.215 & 14.544 & 2.8 & 131.6 & 265.3 & 15.8 & -2.6 \\
2024-12-31 & PS2 & $w$ & 4 &  45 & 15.200 & 14.443 & 2.4 & 139.1 & 256.4 & 15.5 & -2.1 \\
2025-01-07 & PS1 & $w$ & 4 &  45 & 15.188 & 14.378 & 2.2 & 144.4 & 247.5 & 15.2 & -1.7 \\
2025-01-09 & PS2 & $i$ & 4 &  45 & 15.185 & 14.362 & 2.1 & 145.8 & 244.6 & 15.2 & -1.6 \\
2025-01-20 & PS1 & $w$ & 4 &  45 & 15.167 & 14.291 & 1.7 & 152.0 & 224.0 & 14.7 & -0.8 \\
2025-01-29 & PS2 & $w$ & 4 &  45 & 15.152 & 14.260 & 1.6 & 154.1 & 202.6 & 14.3 & -0.2 \\
2025-02-08 & PS1 & $i$ & 4 &  45 & 15.136 & 14.253 & 1.7 & 152.7 & 178.3 & 13.9 & +0.5 \\
2025-02-14 & PS1 & $i$ & 4 &  45 & 15.126 & 14.262 & 1.9 & 150.1 & 165.5 & 13.6 & +0.9 \\
2025-02-14 & PS2 & $i$ & 4 &  45 & 15.126 & 14.262 & 1.9 & 150.1 & 165.5 & 13.6 & +0.9 \\
2025-02-15 & PS1 & $i$ & 4 &  45 & 15.125 & 14.265 & 1.9 & 149.5 & 163.6 & 13.6 & +0.9 \\
2025-02-17 & PS2 & $i$ & 4 &  45 & 15.121 & 14.271 & 2.0 & 148.4 & 160.0 & 13.5 & +1.1 \\
2025-02-22 & PS1 & $w$ & 4 &  45 & 15.113 & 14.291 & 2.1 & 145.1 & 152.0 & 13.3 & +1.4 \\
2025-03-06 & PS1 & $g$ & 1 &  27 & 15.094 & 14.365 & 2.6 & 135.9 & 137.6 & 12.8 & +2.1 \\
2025-03-06 & PS1 & $z$ & 1 &  27 & 15.094 & 14.365 & 2.6 & 135.9 & 137.6 & 12.8 & +2.1 \\
2025-03-16 & PS1 & $r$ & 1 &  27 & 15.079 & 14.452 & 3.0 & 127.5 & 129.1 & 12.4 & +2.7 \\
2025-03-16 & PS1 & $z$ & 1 &  27 & 15.078 & 14.452 & 3.0 & 127.5 & 129.1 & 12.4 & +2.7 \\
2025-03-18 & PS1 & $i$ & 1 &  27 & 15.075 & 14.474 & 3.1 & 125.6 & 127.6 & 12.3 & +2.8 \\
2025-03-18 & PS1 & $r$ & 1 &  27 & 15.075 & 14.474 & 3.1 & 125.6 & 127.6 & 12.3 & +2.8 \\
2025-03-18 & PS2 & $i$ & 4 &  45 & 15.075 & 14.474 & 3.1 & 125.6 & 127.6 & 12.3 & +2.8 \\
\midrule
\multicolumn{12}{p{0.9\textwidth}}{$^{a}$Heliocentric distance.}\\
\multicolumn{12}{p{0.9\textwidth}}{$^{b}$Observer-centric distance.}\\
\multicolumn{12}{p{0.9\textwidth}}{$^{c}$Phase angle.}\\
\multicolumn{12}{p{0.9\textwidth}}{$^{d}$Solar elongation.}\\
\multicolumn{12}{p{0.9\textwidth}}{$^{e}$Position angle of antisolar direction projected in the sky plane of the observer.}\\
\multicolumn{12}{p{0.9\textwidth}}{$^{f}$Position angle of comet's negative heliocentric velocity projected into the sky plane of the observer.}\\
\multicolumn{12}{p{0.9\textwidth}}{$^{g}$Orbital plane angle. Negative values means the observer is below the orbital plane of the comet.}\\
\multicolumn{12}{p{0.9\textwidth}}{$^{\dagger}$Comet blended with a field star, not included for analyses.}\\
\multicolumn{12}{p{0.9\textwidth}}{$^{\ddagger}$Used only for astrometry, because of unremovable strong vignetting from the edge shadow.}\\
\bottomrule

\end{longtable}
}

\section{Results}
\label{sec_rslt}

\subsection{Brightness}
\label{ssec_br}

\begin{figure*}[h!]
\centering
\begin{subfigure}[b]{0.49\textwidth}
\centering
\includegraphics[width=\textwidth]{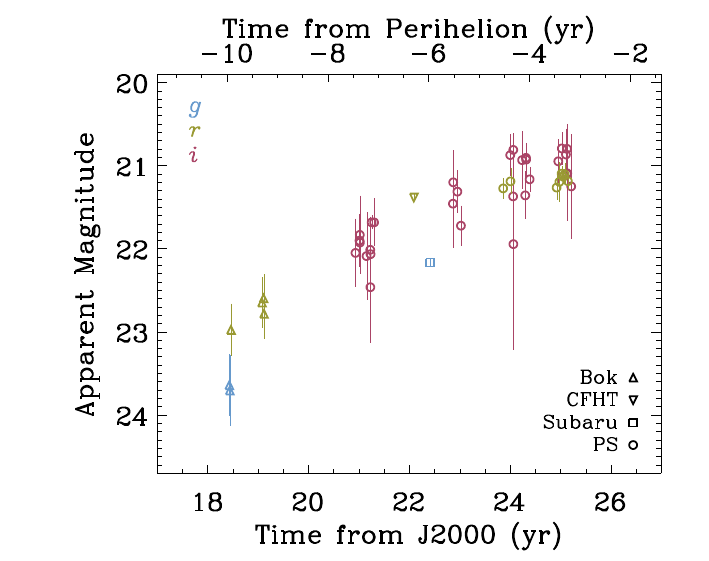}
\caption{
\label{fig:Appmag_vs_t}
}
\end{subfigure}
\begin{subfigure}[b]{0.49\textwidth}
\centering
\includegraphics[width=\textwidth]{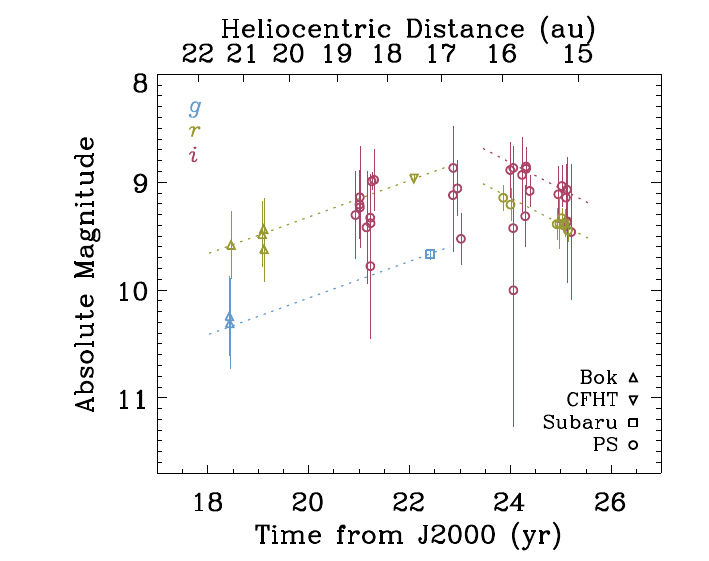}
\caption{
\label{fig:Absmag_vs_t}
}
\end{subfigure}
\caption{
Temporal evolution of (a) apparent magnitude and (b) absolute magnitude of comet C/2025 D1 (Groeller), measured using a $3 \times 10^4$ km radius aperture. The upper axes in panels (a) and (b) mark time with respect to the perihelion passage and heliocentric distance, respectively. Data from different facilities are plotted with distinct symbols, with colours corresponding to the bands to which the photometric reduction was calibrated (see the legends). The dotted lines in panel (b) represent the best-fit linear models to derive the colour indices and activity trends of the comet.
\label{fig:mag_vs_t}
}
\end{figure*}

In Figure \ref{fig:mag_vs_t}(a), we plot the apparent magnitude of comet C/2025 D1, measured using a $3 \times 10^4$ km radius aperture, against time. As the secular apparent lightcurves are highly similar across different apertures, results from other apertures are not included to avoid clutter. We notice that the comet appeared to brighten more rapidly earlier, followed by a slowdown in brightening after approximately late 2023. However, this does not necessarily reflect the intrinsic trend of the comet because of the varying observing geometry. To examine the intrinsic trend, we computed the absolute magnitude, defined as the magnitude of the comet were it at heliocentric and observer-centric distances $r_{\rm H} = \Delta = 1$ au and phase angle $\alpha = 0\degr$, related to the apparent magnitude by
\begin{equation}
    H_{\lambda} = m_{\lambda} - 5 \log \left(r_{\rm H} \Delta \right) - \beta_{\alpha} \alpha
    \label{eq_mabs}.
\end{equation}
\noindent Here, $H_{\lambda}$ and $m_{\lambda}$ are the absolute and apparent magnitudes of the comet, respectively, reduced to bandpass $\lambda$, and we assumed a linear phase function with a phase coefficient of $\beta_{\alpha} = 0.03 \pm 0.01$ mag degree$^{-1}$ typical for comets \citep{1987A&A...187..585M}. The results, shown in Figure \ref{fig:mag_vs_t}(b), reveal distinct brightening trends before and after late 2023 or thereabouts -- the comet intrinsically brightened prior to around late 2023 but began to fade thereafter, despite still approaching perihelion.

To quantify the intrinsic brightening trend, we fitted a linear model to the absolute magnitude of the comet in the time domain using {\tt MPFIT}. Only the $g$- and $r$-band data points at epochs earlier than 2023 were used to best fit the intrinsic brightening trend simultaneously, while the fading trend was fitted with the $r$- and $i$-band data points from late 2023 onward, as the timespan covered by the $i$-band data points is not compatible with the other two bands before 2023. Moreover, including the $i$-band data in the fitting would introduce an extra free parameter in the model, which we prefer not to do, given the quality of these data points. We define the goodness of the fit as
\begin{align}
\begin{split}
\chi^2 \left(\mathbf{\Theta}, C_{1,2} \right) = &
\sum_{i} \left(\frac{H_{1,i} - \mathscr{H}_{1} \left(t_{i}; \mathbf{\Theta} \right)}{\sigma_{1,i}} \right)^{2} + \\
& \sum_{j} \left(\frac{H_{2,j} - \mathscr{H}_{1} \left(t_{j}; \mathbf{\Theta}\right) + C_{1,2}}{\sigma_{2,j}} \right)^{2}
\label{eq_chi2},
\end{split}
\end{align}
\noindent where $\mathscr{H}$ denotes the linear model as a function of ${\bf \Theta}$, the bidimensional parameter (slope and intercept), and time $t$, expressed in years relative to J2000, the subscripts 1 and 2 denote the two simultaneously fitted bandpasses, $C_{1,2} = \mathscr{H}_{1} - \mathscr{H}_{2}$ is the colour index between the two bandpasses, and $\sigma$ is the uncertainty of the absolute magnitude. These parameters were obtained by minimising Equation (\ref{eq_chi2}) using {\tt MPFIT}, i.e.,
\begin{equation}
\frac{\partial \chi^2}{\partial \left(\mathbf{\Theta}, C_{1,2} \right)} = \mathbf{0}
\label{eq_dchi2_to_C}.
\end{equation}
\noindent In Table \ref{tab:bestmdl} we tabulate the best-fit parameters (excluding the intercept, as it is of little interest) for different photometric apertures. The best-fit linear models for the $3 \times 10^4$ km radius aperture are also plotted as dotted lines in Figure \ref{fig:mag_vs_t}(b). Given the uncertainties, the best fits across different photometric apertures are consistent with each other. Therefore, we also present their weighted means and standard deviations in Table \ref{tab:bestmdl}.

\begin{table*}
\caption{Best-fitted Linear Models for the Brightening Trend\label{tab:bestmdl}}
\centering
\begin{tabular}{lclcccc}
\toprule
Time Range & Aperture Radius ($10^4$ km) & \multicolumn{2}{c}{Colour Index $C$} & Slope $\mathcal{S}$ (mag yr$^{-1}$) & $\chi^2$ & Degree of Freedom \\
\midrule
$< 2023$ & 2.5 & $g - r = $ & $+0.75 \pm 0.07$ & $-0.17 \pm 0.04$ & 0.34 & 5 \\
         & 3.0 & & $+0.72 \pm 0.07$ & $-0.15 \pm 0.04$ & 0.46 &\\
         & 3.5 & & $+0.73 \pm 0.07$ & $-0.14 \pm 0.04$ & 0.55 &\\
         & 4.0 & & $+0.76 \pm 0.07$ & $-0.14 \pm 0.03$ & 0.68 &\\
\midrule
$> 2023$ & 2.5 & $r - i = $ & $+0.33 \pm 0.06$ & $+0.24 \pm 0.08$ & 10.01 & 21 \\
         & 3.0 & & $+0.32 \pm 0.06$ & $+0.17 \pm 0.06$ & 14.16 &\\
         & 3.5 & & $+0.37 \pm 0.08$ & $+0.22 \pm 0.08$ & 19.87 &\\
         & 4.0 & & $+0.43 \pm 0.08$ & $+0.24 \pm 0.08$ & 25.18 \\ \hline
\midrule
\multicolumn{2}{l}{Mean $\pm$ standard deviation}
& $g - r =$ & $+0.74 \pm 0.02$ & $-0.15 \pm 0.01$ & -- & --\\
&& $r - i =$ & $+0.35 \pm 0.05$ & $+0.21 \pm 0.04$ & -- & --\\
\bottomrule
\end{tabular}
\tablefoot{
The intercepts of the best-fitted linear models are not presented, as they are of little interest. The reported errors are $1\sigma$ formal uncertainties, properly propagated from the uncertainties in the absolute magnitude, while the errors on the weighted mean values are the weighted standard deviations of the best-fit parameters across multiple apertures. For comparison, the Sun has a colour of $g - r = +0.46 \pm 0.03$ and $r - i = +0.12 \pm 0.03$ \citep{2018ApJS..236...47W}, and the median colour of comets reported by \citet{2012Icar..218..571S} is $g - r = +0.57 \pm 0.05$ and $r - i = +0.22 \pm 0.07$.
}
\end{table*}

\begin{figure*}
\centering
\includegraphics[width=.7\textwidth]{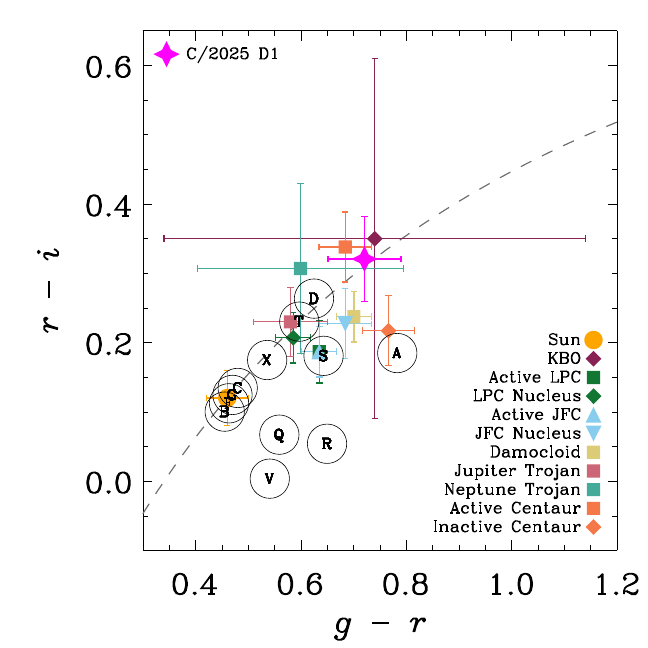}
\caption{The colour of C/2025 D1, measured using a $3 \times 10^4$ km radius photometric aperture, compared with various small solar system body populations \citep[and citations therein]{2003Icar..163..363D, 2007MNRAS.377.1393S, 2012Icar..218..571S, 2015AJ....150..201J, 2023PSJ.....4..135M} and the Sun \citep{2018ApJS..236...47W}. Colours reported in non-SDSS systems have been transformed to the SDSS system following \citet{2006A&A...460..339J}. Open circles with letters at their centres represent main-belt asteroids of specific taxonomic types. Objects with linear reflectivity gradients in the $g - i$ spectral interval form the reddening line, drawn as the grey dashed curve.}
\label{fig:clr}
\end{figure*}

\subsection{Colour}
\label{ssec_clr}

We first examine the derived colours of C/2025 D1. The colour indices $g - r$ and $r - i$ were obtained using observations spanning different time periods, before and after late 2023, respectively. However, \citet{2024PSJ.....5..273H} showed that no long-period comets in their sample exhibited colour variations with heliocentric distance unless at $r_{\rm H} \la 3$ au. Although ultradistant comet C/2010 U3 (Boattini) was reported to display colour variation, this did not occur until the it reached the crystallisation zone at $r_{\rm H} \la 10$ au \citep{2019AJ....157..162H}. We therefore consider our assumption that the colour of C/2025 D1 remains unchanged at the observed heliocentric distances $r_{\rm H} \ga 15$ au to be likely valid. 

Given that the results from different apertures are consistent with each other within their uncertainties, we select those from the $3 \times 10^4$ km radius aperture as representative. Our best-fit colour indices for the comet, $g - r = +0.72 \pm 0.07$ and $r - i = +0.32 \pm 0.06$, are redder than the solar colours \citep[$g - r = +0.46 \pm 0.03$ and $r - i = +0.12 \pm 0.03$;][]{2018ApJS..236...47W} at $3.4\sigma$ and $3.0\sigma$ significance, respectively. Compared to the median colours of comets, reported as $g - r +0.57 \pm 0.05$ and $r - i = +0.22 \pm 0.07$ by \citet{2012Icar..218..571S}, C/2025 D1 is redder at $1.7\sigma$ significance in $g - r$ and $1.1\sigma$ in $r - i$. For a direct comparison, we plot the colour of the comet alongside those of various small solar system body populations and the Sun in Figure \ref{fig:clr}. We can notice that the comet lies closely to the reddening line (corresponding to linear reflectivity gradients) in the colour-colour diagram, as observed for most small bodies in the outer solar system \citep{2002A&A...389..641H}. This in turn lends additional support to our assumption that the comet does not vary in colour. Although C/2025 D1 is redder than typical comets, including long-period comets to which it belongs, it is not the reddest ever measured, given the uncertainties arising from the suboptimal quality of the archival data. The finding that C/2025 D1 is likely not ultrared, despite being at heliocentric distances much larger than some the ultrared Centaurs, may support the hypothesis that the ultrared matter disappears as cometary activity onsets \citep[e.g.,][]{2015AJ....150..201J}.

For completeness, we also calculate the normalised reflectivity gradient \citep{1984AJ.....89..579A, 1986ApJ...310..937J}:
\begin{equation}
S'_{1,2} = 
-\left(\dfrac{2}{\Delta \lambda_{1,2}}\right) \dfrac{10^{0.4 \Delta C_{1,2}} - 1}{10^{0.4 \Delta C_{1,2}} + 1}
\label{eq_Sp},
\end{equation}
\noindent where $\Delta \lambda_{1,2} = \lambda_1 - \lambda_2$ is the central wavelength difference between two bandpasses, and $\Delta C_{1,2}$ is the colour index difference relative to the Sun. The normalisation is chosen at the midpoint of the covered spectral interval. Using the results from the $3 \times 10^4$ km radius aperture, Equation (\ref{eq_Sp}) yields $S'_{g,r} = \left(16 \pm 5 \right) \%$ per $10^3$ \AA~and $S'_{r,i} = \left(14 \pm 5 \right) \%$ per $10^3$ \AA~in the $g - r$ and $r - i$ spectral intervals, respectively. These values are comparable to those of other long-period comets at large heliocentric distances \citep{2018A&A...611A..32K}. A slightly different definition of the normalised reflectivity gradient $\mathcal{S}'_{1,2}$ was introduced by \citet{2002A&A...389..641H}, which is instead normalised at the $V$-band central wavelength. The relationship between the two normalised reflectivity gradients is
\begin{equation}
\mathcal{S}'_{1,2} = \dfrac{S'_{1,2}}{1 + S'_{1,2} \left(\lambda_{V} - \bar{\lambda}_{1,2} \right)}
\label{eq_SpV},
\end{equation}
\noindent where $\bar{\lambda}_{1,2} = \left(\lambda_1 + \lambda_2 \right) / 2$ is the mean wavelength of the spectral interval. The corresponding normalised reflectivity gradients we obtained are $\mathcal{S}'_{g,r} = \left(16 \pm 5\right)\%$ per $10^3$ \AA~in the $g - r$ interval and $\mathcal{S}'_{r,i} = \left(17 \pm 7\right)\%$ per $10^3$ \AA~in the $r - i$ interval, reinforcing that the reflectivity gradient of the comet is statistically linear over the $g - i$ spectral interval.

\subsection{Activity}
\label{ssec_act}

Now we focus on the derived brightening slope of comet C/2025 D1, using results from the $3 \times 10^4$ km radius aperture. According to our best-fit linear models to the available observations, C/2025 D1 intrinsically brightened at a rate of $\mathcal{S} = -0.15 \pm 0.04$ mag yr$^{-1}$ until approximately late 2023 at a heliocentric distance of $r_{\rm H} \approx 16$ au, but subsequently started to fade at a rate of $0.17 \pm 0.06$ mag yr$^{-1}$ towards the present. In the literature, cometary activity is more often characterised by the activity parameter $k$, defined as the slope of the heliocentric magnitude (normalised to observer-centric distance $\Delta = 1$ au and phase angle $\alpha = 0\degr$) with respect to $\log r_{\rm H}$. Instead of refitting the linear model in the $\log r_{\rm H}$ domain, we estimate the activity parameter using the chain rule as
\begin{align}
k & = 5 + \deriv{H}{\left( \log r_{\rm H} \right)}
\nonumber \\
& \approx 5 + \left(\aderiv{H}{t}\right) \frac{r_{\rm H}}{\dot{r}_{\rm H}} \ln 10
\nonumber \\
& \approx 5 - \mathcal{S} \mathcal{T}\ln 10
\label{eq_par_act}.
\end{align}
\noindent Here,
\begin{align}
\mathcal{T} & \triangleq -\frac{r_{\rm H}}{\dot{r}_{\rm H}}
\nonumber \\
& = -\frac{r_{\rm H} }{e \sin \nu}\sqrt{\frac{q \left( 1 + e \right)}{\mu_{\odot}}}
\nonumber \\
& \approx \frac{ r_{\rm H}^2 }{\sqrt{2\mu_{\odot} \left(r_{\rm H} - q \right)}}
\label{eq_t_aux}
\end{align}
\noindent is an auxiliary timescale, simplified by approximating the comet's preperihelion heliocentric trajectory as a parabola in the two-body problem, given eccentricity $e \approx 1$. Here, $q$ and $\nu$ are the perihelion distance and true anomaly of the comet, respectively, and $\mu_{\odot} = 2.96 \times 10^{-4}$ au$^{3}$ d$^{-2}$ is the mass parameter of the Sun. Substitution into Equation (\ref{eq_par_act}) yields $k = 12 \pm 2$ for the brightening phase before approximately late 2023, comparable to values for other long-period comets, whether dynamically new or old, as reported by \citet{2024PSJ.....5..273H} and \citet{2025A&A...697A.210L}. However, from approximately late 2023 onward, the comet faded with an activity parameter of $k = -4 \pm 3$.

Assuming the brightness of the comet is dominated by scattering by dust grains surrounding the nucleus in its coma at such heliocentric distances, we evaluate the evolution of its effective scattering cross-section,
\begin{equation}
\Xi_{\rm e} = \frac{\pi r_{\oplus}^2}{p_r} 10^{0.4 \left(m_{\odot, r} - H_r \right)}
\label{eq_XS},
\end{equation}
\noindent where $p_r = 0.05$ is the nominal $r$-band geometric albedo of cometary dust, $r_{\oplus} = 1$ au, and $m_{\odot, r} = -26.93$ is the apparent $r$-band magnitude of the Sun \citep{2018ApJS..236...47W}, and $H_r$ is the absolute $r$-band magnitude of the comet. Assuming the effective cross-section follows a power-law function of heliocentric distance, i.e., $\Xi_{\rm e} \propto r_{\rm H}^{\gamma}$, we can find the power-law index as
\begin{align}
\gamma & = \dfrac{r_{\rm H}}{\Xi_{\rm e}} \left(\deriv{\Xi_{\rm e}}{r_{\rm H}} \right)
\nonumber \\
& \approx \frac{2}{5}  \mathcal{S} \mathcal{T} \ln 10
\label{eq_slope_xs}.
\end{align}
\noindent Using the $3 \times 10^4$ km radius aperture, Equation (\ref{eq_slope_xs}) yields $\gamma = -2.6 \pm 0.7$ and $4 \pm 1$ for the time periods before and after around late 2023, respectively. The power-law index during the brightening phase is potentially steeper than those measured for ultradistant comets C/2017 K2 (PANSTARRS) and C/2019 E3 (ATLAS) \citep[$\gamma \approx -1$;][]{2021AJ....161..188J, 2024AJ....167..140H}, but the large uncertainty of our results does not preclude the possibility that these comets exhibit a resemblance in their brightening behaviour. Additionally, the power-law index from the brightening period aligns with expectations for CO-driven sublimation activity at similar heliocentric distances \citep[$-2.4 \le \gamma \le -2.2$;][]{2022ApJ...933L..44K}.

We estimate the rate of change in the effective cross-section of the cometary dust by using the power-law index $\gamma$ and again applying the chain rule as
\begin{align}
\dot{\Xi}_{\rm e} & = \deriv{\Xi_{\rm e}}{r_{\rm H}} \dot{r}_{\rm H}
\nonumber \\
& = -\gamma \dfrac{\Xi_{\rm e}}{\mathcal{T}}
\nonumber \\
& \approx -\frac{2}{5} \mathcal{S} \Xi_{\rm e} \ln 10
\label{eq_rate_xs},
\end{align}
\noindent yielding a time-averaged rate of change of $\bar{\dot{\Xi}}_{\rm e} = \left(8 \pm 3 \right) \times 10^2$ km$^{2}$ yr$^{-1}$ for the effective cross-section of dust during the brightening phase of the comet and $\left(-8 \pm 3 \right) \times 10^2$ km$^{2}$ yr$^{-1}$ during the fading phase. Assuming a bulk density of $\rho_{\rm d} = 1$ g cm$^{-3}$ \citep[see][and citations therein]{2024come.book..577E} and a minimum grain radius of 1 mm (see Section \ref{ssec_morph}), we constrain the mass loss of the comet C/2025 D1 to be $\ga\! \left(3 \pm 1 \right) \times 10^4$ g s$^{-1}$ during its intrinsic brightening phase before late 2023. We do not attempt to constrain the mass loss during the fading phase, in that the decrease in the effective cross-section of dust likely resulted from an inefficient supply of newly ejected dust relative to the dust lost from the photometric aperture. 

Based on the available observations, the activity of C/2025 D1 prior to late 2023 was consistent with CO-driven sublimation and did not exhibit peculiar behaviour. However, it subsequently started to fade over a year-long timescale. To our knowledge, no other long-period comet has exhibited similar fading at such large heliocentric distances while on the inbound leg of its orbit. While it is well known that dynamically new comets often exhibit a slowdown in brightening at $r_{\rm H} \approx 3$ au \citep[and citations therein]{2025A&A...697A.210L}, this is inconsistent with our observations for C/2025 D1. 

The decline in intrinsic brightness of the comet resembled that of post-outburst comets. However, the observed fading rate was likely too shallow. For instance, \citet{2022ApJ...933L..44K} reported that ultradistant comet C/2014 UN$_{271}$ (Bernardinelli-Bernstein) faded at a rate of $\sim\!0.01$ mag day$^{-1}$, dominating its post-discovery brightness. Post-outburst Jupiter-family comets 17P/Holmes and 332P/Ikeya-Murakami faded at similar rates of $\la\!0.1$ mag day$^{-1}$ \citep{2014ApJ...787...55I}. Although interstellar object 2I/Borisov exhibited a comparable fading rate of $\bar{\dot{\Xi}}_{\rm e} = -3$ km$^{2}$ day$^{-1}$ on the outbound leg of its orbit \citep[][]{2020ApJ...896L..39J}, the brightening of C/2025 D1 occurred over a timescale orders of magnitude longer than the typical  cometary outbursts, which last no more than a few days. So did the decline of the comet. 

The declining behaviour is nevertheless reminiscent of cometary disintegration. Arguing against this hypothesis, the photocentre of C/2025 D1 remained solid in high signal-to-noise ratio (S/N) observations throughout the fading phase (see Figure \ref{fig:obs}), and its motion is well described by a pure-gravity orbital solution (see Section \ref{ssec_orb}). Both pieces of evidence are inconsistent with a disintegration scenario. 

Thus, we disfavour the hypotheses that C/2025 D1 underwent an outburst or is disintegrating. Instead, we propose that the decrease in the effective cross-section of the cometary dust might result from sublimation of volatiles previously stored in solid state. Applying the simplistic free sublimation model of \citet{1979M&P....21..155C} neglecting heat conduction towards the nucleus interior, with updated physical parameters updated from \citet{2009P&SS...57.2053F}, we find that the onset of CO$_{2}$ would occur at $10 \la r_{\rm H} \la 20$ au. In addition, numerical simulations by \citet{2012AJ....144...97G} suggest crystallisation of amorphous water ice can be triggered at $r_{\rm H} = 16$ au, which is approximately the distance where the fading of C/2025 D1 started. We therefore speculate that the onset of CO$_{2}$ activity and/or crystallisation probably caused the observed fading of the comet starting from around late 2023, whereby the intensified activity in turn reduced the total effective scattering cross-section of cometary dust grains as their CO$_{2}$ and/or H$_{2}$O ices underwent phase transition. This is analogous to the fact that many dynamically new comets slow down in brightening when approaching the Sun within $r_{\rm H} \lesssim 3$ au \citep[e.g.,][]{2025A&A...697A.210L}. The perihelion distance of the comet close to the transition boundary may have amplified the behaviour.

Alternatively, it is possible that the observed fading of C/2025 D1 resulted from the exhaustion of its supervolatiles. Here, we estimate the thickness of its CO ice layer using the same simplistic sublimation model in an order-of-magnitude manner, finding a CO mass flux of $f_{\rm s} \approx 3 \times 10^{-6}$ kg m$^{-2}$ for an isothermal nucleus during the brightening phase of the comet, which lasted for at least $\Delta t \approx 5$ yr. Assuming a bulk density of $\rho_{\rm n} = 0.5$ g cm$^{-3}$ for the cometary nucleus \citep[e.g.,][and citations therein]{2024come.book..249G}, we obtain an erosion of the comet of $f_{\rm s} \Delta t / \rho_{\rm n} \gtrsim\!1$ m in thickness during this period solely due to CO sublimation. Thus, if the CO ice layer was not much thicker, we would expect that the sublimation activity of the comet could be possibly exhausted. Unfortunately, with the available data, we are unable to verify the aforementioned possibilities. Further observations are needed to better understand the specific cause of the observed fading of the comet. We expect that new sky surveys, such as the Vera C. Rubin Observatory Legacy Survey of Space and Time, may discover more distant comets exhibiting similar fading behaviour, if the fading is caused by activity transition and their perihelion distances are barely inside the transition zone. 

Finally, we estimate the nucleus size of C/2025 D1. The brightness of the comet is primarily due to dust particles ejected from the nucleus, hindering effective constraints on the nucleus size using the secular lightcurve. Instead, we still use the simplistic CO sublimation model, assuming a conservative dust-to-gas mass production ratio of 5 following \citet{2019AJ....157...65J}. The model predicts a maximum mass flux of $\sim\!10^{-5}$ kg m$^{-2}$ s$^{-1}$ during the brightening phase of the comet. To sustain the observed activity, a minimum active surface area of $\left( 6 \pm 2 \right) \times 10^5$ m$^{2}$ would be required, corresponding to a lower limit on the radius of a circular active patch of $\left( 4.2 \pm 0.8 \right) \times 10^2$ m. The physical parameters of the comet estimated from other photometric apertures are not statistically different from those obtained with the $3 \times 10^4$ km radius aperture. Our constraint on the nucleus size here is likely no better than an order-of-magnitude estimate, because the model oversimplifies the actual complexity of the cometary activity in C/2025 D1.

\begin{figure*}
\centering
\begin{subfigure}[b]{0.49\textwidth}
\centering
\includegraphics[width=\textwidth]{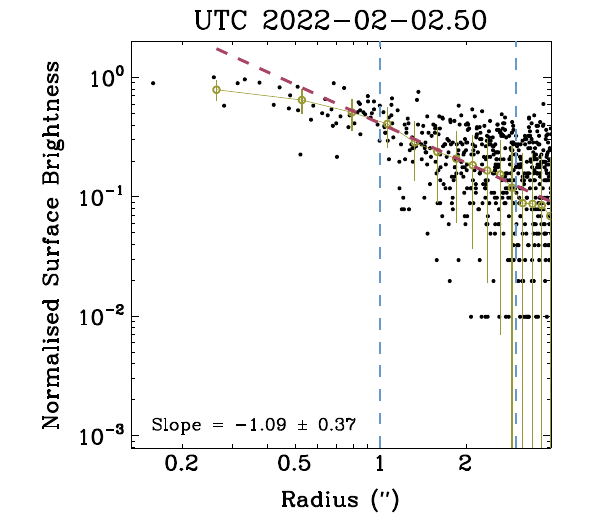}
\caption{
\label{fig:rprof_CFH}
}
\end{subfigure}
\begin{subfigure}[b]{0.49\textwidth}
\centering
\includegraphics[width=\textwidth]{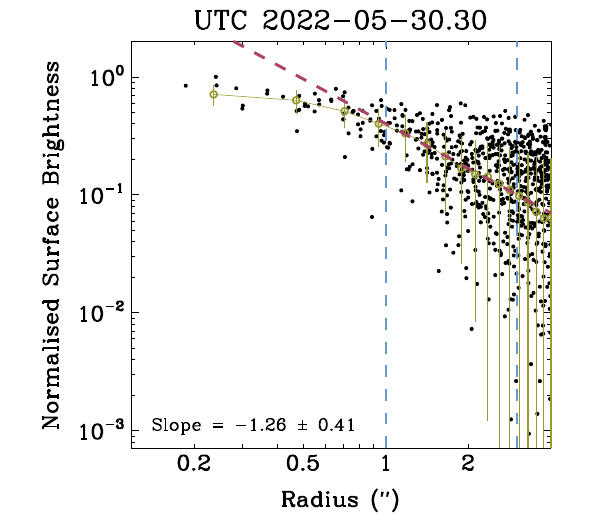}
\caption{
\label{fig:rprof_HSC}
}
\end{subfigure}
\caption{
Radial surface brightness profiles of comet C/2025 D1 on (a) 2022 February 2 from CFHT and (b) 2022 May 30 from HSC. These images have higher S/N for the comet than any other observations we collected. The azimuthally averaged surface brightness is plotted in olive. The best-fit radial surface brightness model is shown as a red dashed line, with the derived logarithmic surface brightness slope indicated in the lower left corner. The boundaries of the fitted region are marked by the two vertical blue dashed lines.
\label{fig:radprof}
}
\end{figure*}

\subsection{Morphology}
\label{ssec_morph}

The physical properties of the dust environment of C/2025 D1 can be inferred from its morphology, which is dominated by ejected dust grains. Unfortunately, the faintness of the comet precluded effective constraints on its dust environment through modelling of the dust morphology \citep[e.g.,][]{2022Univ....8..366M}. Visual inspection reveals that the comet maintained a largely circularly symmetric appearance throughout the observed period, irrespective of the orbital plane angle (see Figure \ref{fig:obs} and Table \ref{tab:vgeo}). Notably, the CFHT observation was acquired when Earth happened to be in the orbital plane of the comet, a configuration in which many comets exhibit a linear morphology centred on the projected orbital plane. This is because the in-orbit dispersion of small dust particles (primarily caused by solar radiation pressure) is much greater than the out-of-plane dispersion (primarily due to the initial dust ejection speed). The fact that the comet did not exhibit an increasingly diffuse appearance in observations from the fading phase further argues against the disintegration possibility. The morphology of C/2025 D1 closely resembles that of ultradistant comets C/2017 K2 (PANSTARRS) and C/2019 E3 (ATLAS), whose comae were reported to be optically dominated by large dust grains \citep[at least submillimetre to millimetre scaled;][]{2024AJ....167..140H,2019AJ....157...65J}. Accordingly, we postulate that the dust environment of C/2025 D1 is similarly dominated by large dust particles, as a possible consequence of supervolatile sublimation at large heliocentric distances. Future observations are needed to investigate the physical parameters of the dust environment of the comet.

We attempted to infer the activity mechanism of the comet from its surface brightness profile. The highest S/N archival observations of the comet were obtained on 2022 February 2 from CFHT and May 30 from HSC. To compute the surface brightness profile, we used a series of circular apertures with radii ranging from 1 pixel to 4\arcsec~in increments of 1 pixel, centred on the photocentre of the comet, to obtain the enclosed total brightness. The sky background and associated uncertainty were determined as described in Section \ref{sec_obs}. The surface brightness was then calculated by differentiating the total brightness with respect to the aperture area. We plot the surface brightness profiles of the comet as functions of radial distance from the photocentre of the comet for the aforementioned two epochs in Figure \ref{fig:radprof}, normalised to the peak surface brightness. As evident, the low S/N causes the signal of the coma quickly fade into the sky background beyond $\ga\!3\arcsec$ from the photocentre. However, the profile within $\sim\!1\arcsec$ from the photocentre are distorted by convolution with the seeing and optical system. Therefore, we fitted a power-law model using {\tt MPFIT} to the profiles in an annulus from 1\arcsec~to 3\arcsec~of the photocentre, which we consider more representative of the true profile. This yields the logarithmic surface brightness slope, which is the power-law index of the model, to be $-1.1 \pm 0.4$ and $-1.3 \pm 0.4$ for the CFHT and HSC observations, respectively, both in agreement with a coma in steady state given the uncertainties. These results lend support to the interpretation that the observed activity of the comet is driven by sublimation of supervolatiles, possibly CO and/or CO$_{2}$. 

\begin{table*}
\caption{Best-fit Orbital Solution for C/2025 D1 (Groeller)
\label{tab:orb}}
\centering
\begin{tabular}{lcl}
\toprule
\multicolumn{2}{c}{Quantity} & Value \\
\midrule
Perihelion distance (au) & $q$
       & 14.119662(58) \\
Eccentricity & $e$
       & 1.003053(12) \\ 
Inclination (\degr) & $i$
       & 84.466759(26) \\ 
Argument of perihelion (\degr) & $\omega$
                 & 185.88487(38) \\ 
Longitude of ascending node (\degr) & ${\Omega}$
                 & 312.897549(19) \\ 
Time of perihelion passage (TDB)\tablefootmark{$\dagger$} & $t_{\rm p}$
                  & 2028 May 19.789(12) \\
\midrule
\multicolumn{2}{l}{Number of observations used (discarded)} & 122 (28) \\
\multicolumn{2}{l}{Observed arc} & 2018 Jun 6-2025 May 2 \\
\multicolumn{2}{l}{Residual rms (\arcsec)} & 0.435 \\
\multicolumn{2}{l}{Normalised residual rms} & 1.038 \\
\bottomrule
\multicolumn{3}{l}{$^{\dagger}$The uncertainty is in days.}
\end{tabular}
\tablefoot{
The Keplerian orbital elements are referred to the heliocentric J2000 ecliptic reference frame at an osculating epoch of TDB 2025 May 2.0, with $1\sigma$ formal uncertainties expressed in the parenthesis notation.
}
\end{table*}

\subsection{Orbit}
\label{ssec_orb}

In addition to our own astrometric measurements, we obtained more recent astrometry of comet C/2025 D1 using the MPC Observations API tool\footnote{\url{https://minorplanetcenter.net/mpcops/documentation/observations-api/}}. The observations were debiased according to star catalogues following \citet{2020Icar..33913596E}, weighted using the measurement uncertainties or following \citet{2017Icar..296..139V} for MPC observations without reported uncertainties, and best fitted with a gravity-only model (Keplerian orbital elements as the six free parameters) in {\tt Find\_Orb}\footnote{\url{https://www.projectpluto.com/find_orb.htm}}, which utilised the planetary and lunar ephemeris DE441 \citep{2021AJ....161..105P} and accounted for gravitational perturbations from the eight major planets, the Moon, Pluto, and the 16 most massive main-belt asteroids. For completeness, the oblateness of the Sun and the Earth, as well as post-Newtonian corrections, were incorporated, although they had negligible effects on the orbital determination. The preliminary gravity-only solution yielded no noticeable systematic trends in the observed-minus-calculated ($O - C$) residuals but showed outliers beyond $3\sigma$. We rejected these measurements, including 16 of the 81 PS measurements and 12 of the 59 measurements downloaded from the MPC. The rejected PS measurements are all from extremely low S/N images. We performed a test by including the rejected PS data points and inflating their uncertainties to a common value of 0\farcs5~based on their $O-C$ residuals, rather than discarding them. The resulting gravity-only orbital solution remained statistically consistent, indicating the robustness of the solution. Table \ref{tab:orb} lists the best-fit Keplerian orbital elements of the comet from the version without error inflation. 

Although the best-fit osculating heliocentric orbit of comet C/2025 D1 at the referenced epoch is slightly hyperbolic, this does not necessarily imply an extrasolar origin. We studied the orbital evolution of comet C/2025 D1 by integrating the nominal orbit and 1000 Monte Carlo massless clones, generated according to the nominal orbit and its full covariance matrix, backward in time. The covariance matrix was properly propagated from the uncertainties in the astrometric observations. While time consuming, the Monte Carlo approach provides a straightforward and reliable method to map uncertainties. We first derived the ``original'' orbit of the comet, defined as its solar system barycentric osculating orbit at a preperihelion heliocentric distance of $r_{\rm H} = 250$ au \citep{2001A&A...375..643D}. Since this moment falls within the time coverage of DE441, we still employed {\tt Find\_Orb} with the same configuration as in the orbit determination to perform the N-body integration for the 1001 test particles. We recorded the epochs when the test particles reached preperihelion $r_{\rm H} = 250$ au and obtained their corresponding heliocentric states through interpolation with relevant tools by JPL's Navigation and Ancillary Information Facility \citep[NAIF;][]{1996P&SS...44...65A,2018P&SS..150....9A}. We then computed the solar system barycentric states by shifting the origin from the Sun to the solar system barycentre with DE441 and converted the results to Keplerian orbital elements. The distribution of each orbital element was well described by a Gaussian, and therefore, we calculated the mean and standard deviation, as shown in Table \ref{tab:orb_OF}. The solar system barycentric eccentricity $e < 1$ and semimajor axis $a = \left(3.6 \pm 0.1 \right) \times 10^4$ au unequivocally indicate that C/2025 D1 is a long-period comet from the Oort Cloud on the outskirts of our solar system. Using Kepler's third law, we estimate that the previous apparition of the comet occurred $\sim\!7$ Myr ago.

\begin{center}
\begin{figure*}
\includegraphics[width=1\textwidth]{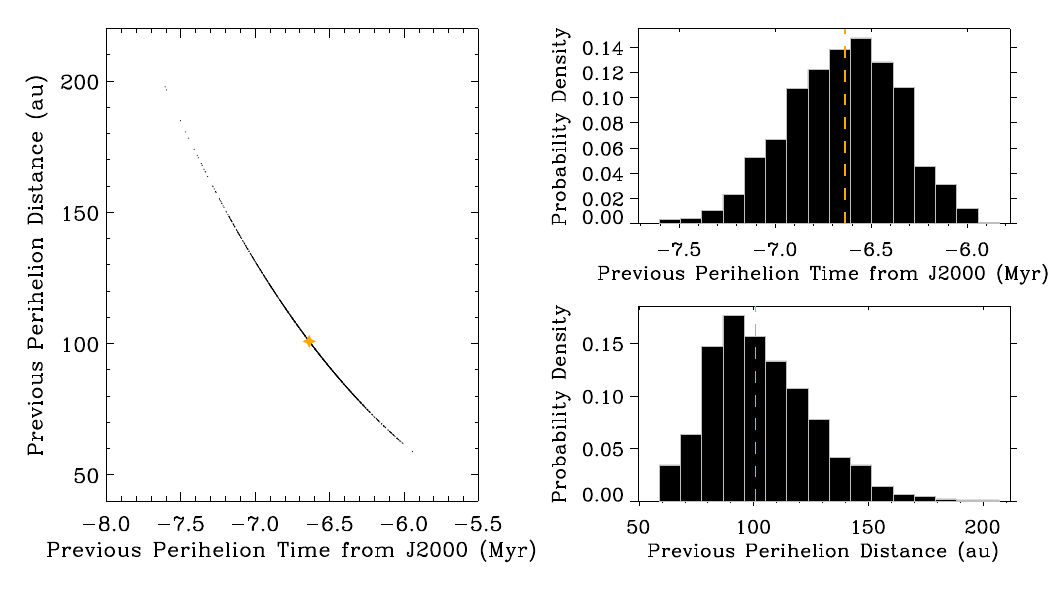}
\caption{Orbit of C/2025 D1 at its previous perihelion, derived from the Monte Carlo clones. The left panel shows the orbital uncertainty in the space of perihelion time and distance, with the corresponding probability density distributions for each element presented in the right two panels. The nominal orbit is highlighted in orange. For the N-body integration, we adopted a local mass density in the solar neighbourhood of $\rho_{\ast} = 0.185~M_{\odot} $ pc$^{-3}$ \citep{1984ApJ...276..169B}.}
\label{fig:peri_prev}
\end{figure*}
\end{center}

To investigate the previous return, we must take into account at least the influence of the Galactic tide, which becomes pronounced at the Oort Cloud heliocentric distances \citep[e.g.,][]{1986Icar...65...13H}. However, this effect is not included in {\tt Find\_Orb}, and the previous perihelion occurs well beyond the time coverage of DE441. We therefore employed {\tt mercury6}, modified from the original version by \citet{1999MNRAS.304..793C} to include the Galactic tide and post-Newtonian corrections, to study the previous perihelion return of C/2025 D1 using the same 1001 test particles. For consistency, we used the states of planetary systems from DE441 and the best-fit orbital elements of C/2025 D1 as initial conditions for the N-body integration in {\tt mercury6}. The same gravitational perturbations were included, although the Earth-Moon system was represented by their barycentre rather than separately as in the orbit determination. To model the Galactic tide, we adopted a nominal mass density in the solar neighbourhood of $\rho_{\ast} = 0.185~M_{\odot}$ pc$^{-3}$ \citep[$M_{\odot}$ is the solar mass;][]{1984ApJ...276..169B} in the integration. Using pertinent NAIF routines, we monitored and recorded the perihelion passages of the nominal orbit and its Monte Carlo clones by searching for local minima in their heliocentric distances prior to the current apparition. Test particles reaching beyond $r_{\rm H} = 10^6$ au were considered lost from the solar system, and 10 such test particles were dropped from our backward integration. Our results for the remaining 990 clones and the nominal orbit reveal that the previous perihelion of C/2025 D1 occurred between $\sim\!5.9$ and 7.6 Myr ago, at a heliocentric distance between $\sim\!60$ and 200 au (see Figure \ref{fig:peri_prev}), suggesting that the comet is dynamically new. We acknowledge the limitation that our N-body integration did not incorporate perturbations from random passing stars, which can influence the motion of long-period comets. Such close stellar encounters are possibly not rare, occurring at a rate of $\sim\!20$ Myr$^{-1}$ within 1 pc of the Sun \citep{2018A&A...616A..37B}. Subsequent to our numerical integration, \citet{2025arXiv250820780D} updated the Catalogue of Cometary Orbits and their Dynamical Evolution (CODE catalogue), including C/2025 D1. While their result without stellar perturbations yields a previous perihelion distance of the comet largely consistent with ours, the model taking into account stellar perturbations suggests a much further perihelion distance of $q \ga 10^3$ au. Therefore, we are confident that C/2025 D1 is a dynamically new comet.

We also studied the future orbital evolution of the comet. Using {\tt Find\_Orb}, we first determined the ``future'' orbit at a postperihelion heliocentric distance of $r_{\rm H} = 250$ au for the nominal orbit and its 1000 Monte Carlo clones. Table \ref{tab:orb_OF} presents the mean orbital elements and their uncertainties of the future orbit, as their uncertainty distributions closely follow Gaussian distributions. Our findings indicate that, due to planetary perturbations, the future orbit of the comet possibly becomes hyperbolic, implying its escape from the solar system. Although this result is not statistically significant ($1.8\sigma$), we argue that the comet will likely be lost from the solar system, in that the 46 test particles with solar system barycentric eccentricity $<1$ (a 4.6\% fraction of the total) all have semimajor axes $\ga\!10^6$ au for their future orbits, comparable to the distances of the nearest stars to the solar system. This finding aligns with the predictions in the CODE catalogue by \citet{2025arXiv250820780D}, which incorporates the Galactic tide (and stellar perturbations) and estimates a 0.1\%~probability of the comet remaining bound to the solar system after the current perihelion passage.

We acknowledge that the above analysis does not account for nongravitational effects, which are commonly observed in comets due to anisotropic activity and/or splitting \citep[e.g.,][]{2004come.book..301B,2004come.book..137Y}. To investigate whether the observed decline in intrinsic brightness of comet C/2025 D1 could result from disintegration, we refitted the astrometric measurements in {\tt Find\_Orb} using a nongravitational model that included the area-to-mass ratio (AMR) as an additional free parameter to account for solar radiation pressure. Our findings indicate that the best-fit AMR, $\left( 6 \pm 9 \right) \times 10^{-2}$ m$^{2}$ kg$^{-1}$, is not statistically significant and that the normalised residual rms showed negligible improvement. Therefore, given also the multi-year observed arc, we conclude with confidence that there is no astrometric evidence of disintegration in comet C/2025 D1, and that the gravity-only analysis remains robust.

\begin{table*}
\caption{Original and Future Orbits of C/2025 D1 (Groeller)
\label{tab:orb_OF}}
\centering
\begin{tabular}{lccc}
\toprule
\multicolumn{2}{c}{Quantity} & Original & Future \\
\midrule
Pericentric distance (au) & $q$
       & 14.108802(53)
       & 14.108838(68) \\ 
Eccentricity & $e$
       & 0.999607(12)
       & 1.000022(12) \\ 
Reciprocal of semimajor axis ($10^{-5}$ au$^{-1}$) & $a^{-1}$
       & 2.783(87)
       & -0.155(88) \\
Inclination (\degr) & $i$
       & 84.505187(27)
       & 84.496934(75) \\ 
Argument of periapsis (\degr) & $\omega$
                 & 186.05861(38)
                 & 186.04463(37) \\ 
Longitude of ascending node (\degr) & ${\Omega}$
                 & 312.910794(20)
                 & 312.908494(34) \\ 
Time of periapsis (TDB)\tablefootmark{$\dagger$} & $t_\mathrm{p}$
                  & 2028 May 25.831(13)
                  & 2028 May 25.8670(82) \\
\midrule
\multicolumn{2}{l}{Epoch (TDB)\tablefootmark{$\dagger$}} & 
    1707 Aug $23.4 \pm 4.1$ & 
    2348 Oct $8.6 \pm 4.1$ \\
\bottomrule
\multicolumn{4}{l}{$^{\dagger}$The uncertainties are in days.}
\end{tabular}
\tablefoot{
The Keplerian orbital elements of the original and future orbits are both referred to the solar system barycentric J2000 ecliptic reference frame at the osculating epochs when the comet is at pre- and post-perihelion heliocentric distances $r_{\rm H} = 250$ au, respectively. The values and their uncertainties (indicated in parentheses) were computed from the means and standard deviations of the 1001 MC clones.}
\end{table*}

\section{Summary}
\label{sec_sum}

In this paper, we presented a study of long-period comet C/2025 D1 (Groeller), whose perihelion distance is larger than any other known comets, using archival observations. The key findings are as follows:
\begin{enumerate}
    \item The ultradistant comet exhibited preperihelion activity at heliocentric distances $r_{\rm H} \ga 20$ au. While its activity intensified, resembling other long-period comets at earlier epochs, it began to decline starting approximately from late 2023 at heliocentric distance $r_{\rm H} \approx 16$ au. The mechanism for the fading is unclear, but we disfavour an outburst and disintegration hypotheses. Instead, we conjecture that it may be related to the onset of CO$_{2}$ activity and/or crystallisation of amorphous ice, or that the comet might have exhausted its supervolatile supply.

    \item Assuming the activity trend is not bandpass dependent, we derived the colour of the comet to be $g - r = +0.72 \pm 0.07$ and $r - i = +0.32 \pm 0.06$ (measured from a $3 \times 10^4$ km radius aperture). This colour is significantly redder than the Sun and also redder than many solar system comets. However, it is likely not the reddest ever measured.

    \item By measuring the astrometry of the comet in archival data, we improved the orbital solution, based on which we derived the original and future orbits of the comet in a Monte Carlo approach. Our N-body numerical integration incorporating the influence from the Galactic tide reveals that the comet is highly likely dynamically new from the Oort Cloud. The previous perihelion occurred $\ga\!6$ Myr ago at a distance of $\ga\!60$ au from the Sun. However, subsequent to the current apparition, it is highly likely that the comet will become gravitationally unbound to the solar system due to planetary perturbations.

    \item From the best archival observations, we computed its surface brightness profiles and found them to be consistent with a coma in steady state, which implies the observed activity being driven by sublimation of supervolatiles. The coma remained largely circularly symmetric throughout the observed period despite of changes in the orbital plane angle. Therefore, we postulate that the dust environment of the comet primarily consisted of large dust grains (at least submillimetre to millimetre scaled).

    \item Our model-dependent estimate, assuming CO-driven activity, constrains the nucleus radius to be $\ga\!0.4$ km in order to sustain the observed brightening trend in the brightening phase prior to approximately late 2023.
\end{enumerate}

\begin{acknowledgements}
We thank Olivier Hainaut for a thorough and detailed review on our manuscript, Bill Gray for implementing his orbit determination software {\tt Find\_Orb}, Xiang Tang for setting up a server at the Shanghai Astronomical Observatory for the N-body integration runs, and observers who measured and submitted their astrometric observations of C/2025 D1 (Groeller) to the Minor Planet Center.

Pan-STARRS is supported by the National Aeronautics and Space Administration under Grants 80NSSC18K0971 and 80NSSC21K1572 issued through the SSO Near Earth Object Observations Program.

This research has made use of data and services provided by the International Astronomical Union's Minor Planet Center and the facilities of the Canadian Astronomy Data Centre operated by the National Research Council of Canada with the support of the Canadian Space Agency. This work was financially supported by a grant to M.T.H. J.C.S. and X.S. are financially supported by the National Natural Science Foundation of China through Grants Nos. 12173093 and 12233003, respectively.

\end{acknowledgements}

\bibliography{2025D1}{}

\begin{thebibliography}{63}
\expandafter\ifx\csname natexlab\endcsname\relax\def\natexlab#1{#1}\fi

\bibitem[{{Acton} {et~al.}(2018){Acton}, {Bachman}, {Semenov}, \&
  {Wright}}]{2018P&SS..150....9A}
{Acton}, C., {Bachman}, N., {Semenov}, B., \& {Wright}, E. 2018, \planss, 150,
  9

\bibitem[{{Acton}(1996)}]{1996P&SS...44...65A}
{Acton}, C.~H. 1996, \planss, 44, 65

\bibitem[{{A'Hearn} {et~al.}(1984){A'Hearn}, {Schleicher}, {Millis}, {Feldman},
  \& {Thompson}}]{1984AJ.....89..579A}
{A'Hearn}, M.~F., {Schleicher}, D.~G., {Millis}, R.~L., {Feldman}, P.~D., \&
  {Thompson}, D.~T. 1984, \aj, 89, 579

\bibitem[{{Aune} {et~al.}(2003){Aune}, {Boulade}, {Charlot}, {Abbon},
  {Borgeaud}, {Carton}, {Carty}, {Da Costa}, {Desforge}, {Deschamps},
  {Eppell{\'e}}, {Gallais}, {Gosset}, {Granelli}, {Gros}, {de Kat}, {Loiseau},
  {Ritou}, {Rouss{\'e}}, {Starzynski}, {Vignal}, \&
  {Vigroux}}]{2003SPIE.4841..513A}
{Aune}, S., {Boulade}, O., {Charlot}, X., {et~al.} 2003, in Society of
  Photo-Optical Instrumentation Engineers (SPIE) Conference Series, Vol. 4841,
  Instrument Design and Performance for Optical/Infrared Ground-based
  Telescopes, ed. M.~{Iye} \& A.~F.~M. {Moorwood}, 513--524

\bibitem[{{Bahcall}(1984)}]{1984ApJ...276..169B}
{Bahcall}, J.~N. 1984, \apj, 276, 169

\bibitem[{{Bailer-Jones} {et~al.}(2018){Bailer-Jones}, {Rybizki}, {Andrae}, \&
  {Fouesneau}}]{2018A&A...616A..37B}
{Bailer-Jones}, C.~A.~L., {Rybizki}, J., {Andrae}, R., \& {Fouesneau}, M. 2018,
  \aap, 616, A37

\bibitem[{{Belousov} \& {Pavlov}(2024)}]{2024Icar..41516066B}
{Belousov}, D.~V. \& {Pavlov}, A.~K. 2024, \icarus, 415, 116066

\bibitem[{{Boehnhardt}(2004)}]{2004come.book..301B}
{Boehnhardt}, H. 2004, in Comets II, ed. M.~C. {Festou}, H.~U. {Keller}, \&
  H.~A. {Weaver}, 301

\bibitem[{{Chambers}(1999)}]{1999MNRAS.304..793C}
{Chambers}, J.~E. 1999, \mnras, 304, 793

\bibitem[{{Chambers} {et~al.}(2016){Chambers}, {Magnier}, {Metcalfe},
  {Flewelling}, {Huber}, {Waters}, {Denneau}, {Draper}, {Farrow}, {Finkbeiner},
  {Holmberg}, {Koppenhoefer}, {Price}, {Rest}, {Saglia}, {Schlafly}, {Smartt},
  {Sweeney}, {Wainscoat}, {Burgett}, {Chastel}, {Grav}, {Heasley}, {Hodapp},
  {Jedicke}, {Kaiser}, {Kudritzki}, {Luppino}, {Lupton}, {Monet}, {Morgan},
  {Onaka}, {Shiao}, {Stubbs}, {Tonry}, {White}, {Ba{\~n}ados}, {Bell},
  {Bender}, {Bernard}, {Boegner}, {Boffi}, {Botticella}, {Calamida},
  {Casertano}, {Chen}, {Chen}, {Cole}, {Deacon}, {Frenk}, {Fitzsimmons},
  {Gezari}, {Gibbs}, {Goessl}, {Goggia}, {Gourgue}, {Goldman}, {Grant},
  {Grebel}, {Hambly}, {Hasinger}, {Heavens}, {Heckman}, {Henderson}, {Henning},
  {Holman}, {Hopp}, {Ip}, {Isani}, {Jackson}, {Keyes}, {Koekemoer}, {Kotak},
  {Le}, {Liska}, {Long}, {Lucey}, {Liu}, {Martin}, {Masci}, {McLean}, {Mindel},
  {Misra}, {Morganson}, {Murphy}, {Obaika}, {Narayan}, {Nieto-Santisteban},
  {Norberg}, {Peacock}, {Pier}, {Postman}, {Primak}, {Rae}, {Rai}, {Riess},
  {Riffeser}, {Rix}, {R{\"o}ser}, {Russel}, {Rutz}, {Schilbach}, {Schultz},
  {Scolnic}, {Strolger}, {Szalay}, {Seitz}, {Small}, {Smith}, {Soderblom},
  {Taylor}, {Thomson}, {Taylor}, {Thakar}, {Thiel}, {Thilker}, {Unger},
  {Urata}, {Valenti}, {Wagner}, {Walder}, {Walter}, {Watters}, {Werner},
  {Wood-Vasey}, \& {Wyse}}]{2016arXiv161205560C}
{Chambers}, K.~C., {Magnier}, E.~A., {Metcalfe}, N., {et~al.} 2016, arXiv
  e-prints, arXiv:1612.05560

\bibitem[{{Cowan} \& {A'Hearn}(1979)}]{1979M&P....21..155C}
{Cowan}, J.~J. \& {A'Hearn}, M.~F. 1979, Moon and Planets, 21, 155

\bibitem[{{Dandy} {et~al.}(2003){Dandy}, {Fitzsimmons}, \&
  {Collander-Brown}}]{2003Icar..163..363D}
{Dandy}, C.~L., {Fitzsimmons}, A., \& {Collander-Brown}, S.~J. 2003, \icarus,
  163, 363

\bibitem[{{Dybczy{\'n}ski}(2001)}]{2001A&A...375..643D}
{Dybczy{\'n}ski}, P.~A. 2001, \aap, 375, 643

\bibitem[{{Dybczy{\'n}ski} \& {Kr{\'o}likowska}(2025)}]{2025arXiv250820780D}
{Dybczy{\'n}ski}, P.~A. \& {Kr{\'o}likowska}, M. 2025, arXiv e-prints,
  arXiv:2508.20780

\bibitem[{{Eggl} {et~al.}(2020){Eggl}, {Farnocchia}, {Chamberlin}, \&
  {Chesley}}]{2020Icar..33913596E}
{Eggl}, S., {Farnocchia}, D., {Chamberlin}, A.~B., \& {Chesley}, S.~R. 2020,
  \icarus, 339, 113596

\bibitem[{{Engrand} {et~al.}(2024){Engrand}, {Lasue}, {Wooden}, \&
  {Zolensky}}]{2024come.book..577E}
{Engrand}, C., {Lasue}, J., {Wooden}, D.~H., \& {Zolensky}, M.~E. 2024, in
  Comets III, ed. K.~J. {Meech}, M.~R. {Combi}, D.~{Bockel{\'e}e-Morvan}, S.~N.
  {Raymodn}, \& M.~E. {Zolensky}, 577--620

\bibitem[{{Fray} \& {Schmitt}(2009)}]{2009P&SS...57.2053F}
{Fray}, N. \& {Schmitt}, B. 2009, \planss, 57, 2053

\bibitem[{{Gaia Collaboration} {et~al.}(2018){Gaia Collaboration}, {Brown},
  {Vallenari}, {Prusti}, {de Bruijne}, {Babusiaux}, {Bailer-Jones}, {Biermann},
  {Evans}, {Eyer}, {Jansen}, {Jordi}, {Klioner}, {Lammers}, {Lindegren},
  {Luri}, {Mignard}, {Panem}, {Pourbaix}, {Randich}, {Sartoretti}, {Siddiqui},
  {Soubiran}, {van Leeuwen}, {Walton}, {Arenou}, {Bastian}, {Cropper},
  {Drimmel}, {Katz}, {Lattanzi}, {Bakker}, {Cacciari}, {Casta{\~n}eda},
  {Chaoul}, {Cheek}, {De Angeli}, {Fabricius}, {Guerra}, {Holl}, {Masana},
  {Messineo}, {Mowlavi}, {Nienartowicz}, {Panuzzo}, {Portell}, {Riello},
  {Seabroke}, {Tanga}, {Th{\'e}venin}, {Gracia-Abril}, {Comoretto},
  {Garcia-Reinaldos}, {Teyssier}, {Altmann}, {Andrae}, {Audard},
  {Bellas-Velidis}, {Benson}, {Berthier}, {Blomme}, {Burgess}, {Busso},
  {Carry}, {Cellino}, {Clementini}, {Clotet}, {Creevey}, {Davidson}, {De
  Ridder}, {Delchambre}, {Dell'Oro}, {Ducourant},
  {Fern{\'a}ndez-Hern{\'a}ndez}, {Fouesneau}, {Fr{\'e}mat}, {Galluccio},
  {Garc{\'\i}a-Torres}, {Gonz{\'a}lez-N{\'u}{\~n}ez}, {Gonz{\'a}lez-Vidal},
  {Gosset}, {Guy}, {Halbwachs}, {Hambly}, {Harrison}, {Hern{\'a}ndez},
  {Hestroffer}, {Hodgkin}, {Hutton}, {Jasniewicz}, {Jean-Antoine-Piccolo},
  {Jordan}, {Korn}, {Krone-Martins}, {Lanzafame}, {Lebzelter}, {L{\"o}ffler},
  {Manteiga}, {Marrese}, {Mart{\'\i}n-Fleitas}, {Moitinho}, {Mora}, {Muinonen},
  {Osinde}, {Pancino}, {Pauwels}, {Petit}, {Recio-Blanco}, {Richards},
  {Rimoldini}, {Robin}, {Sarro}, {Siopis}, {Smith}, {Sozzetti}, {S{\"u}veges},
  {Torra}, {van Reeven}, {Abbas}, {Abreu Aramburu}, {Accart}, {Aerts},
  {Altavilla}, {{\'A}lvarez}, {Alvarez}, {Alves}, {Anderson}, {Andrei},
  {Anglada Varela}, {Antiche}, {Antoja}, {Arcay}, {Astraatmadja}, {Bach},
  {Baker}, {Balaguer-N{\'u}{\~n}ez}, {Balm}, {Barache}, {Barata}, {Barbato},
  {Barblan}, {Barklem}, {Barrado}, {Barros}, {Barstow}, {Bartholom{\'e}
  Mu{\~n}oz}, {Bassilana}, {Becciani}, {Bellazzini}, {Berihuete}, {Bertone},
  {Bianchi}, {Bienaym{\'e}}, {Blanco-Cuaresma}, {Boch}, {Boeche}, {Bombrun},
  {Borrachero}, {Bossini}, {Bouquillon}, {Bourda}, {Bragaglia}, {Bramante},
  {Breddels}, {Bressan}, {Brouillet}, {Br{\"u}semeister}, {Brugaletta},
  {Bucciarelli}, {Burlacu}, {Busonero}, {Butkevich}, {Buzzi}, {Caffau},
  {Cancelliere}, {Cannizzaro}, {Cantat-Gaudin}, {Carballo}, {Carlucci},
  {Carrasco}, {Casamiquela}, {Castellani}, {Castro-Ginard}, {Charlot},
  {Chemin}, {Chiavassa}, {Cocozza}, {Costigan}, {Cowell}, {Crifo}, {Crosta},
  {Crowley}, {Cuypers}, {Dafonte}, {Damerdji}, {Dapergolas}, {David}, {David},
  {de Laverny}, \& {De Luise}}]{2018A&A...616A...1G}
{Gaia Collaboration}, {Brown}, A.~G.~A., {Vallenari}, A., {et~al.} 2018, \aap,
  616, A1

\bibitem[{{Gaia Collaboration} {et~al.}(2023){Gaia Collaboration}, {Vallenari},
  {Brown}, {Prusti}, {de Bruijne}, {Arenou}, {Babusiaux}, {Biermann},
  {Creevey}, {Ducourant}, {Evans}, {Eyer}, {Guerra}, {Hutton}, {Jordi},
  {Klioner}, {Lammers}, {Lindegren}, {Luri}, {Mignard}, {Panem}, {Pourbaix},
  {Randich}, {Sartoretti}, {Soubiran}, {Tanga}, {Walton}, {Bailer-Jones},
  {Bastian}, {Drimmel}, {Jansen}, {Katz}, {Lattanzi}, {van Leeuwen}, {Bakker},
  {Cacciari}, {Casta{\~n}eda}, {De Angeli}, {Fabricius}, {Fouesneau},
  {Fr{\'e}mat}, {Galluccio}, {Guerrier}, {Heiter}, {Masana}, {Messineo},
  {Mowlavi}, {Nicolas}, {Nienartowicz}, {Pailler}, {Panuzzo}, {Riclet}, {Roux},
  {Seabroke}, {Sordo}, {Th{\'e}venin}, {Gracia-Abril}, {Portell}, {Teyssier},
  {Altmann}, {Andrae}, {Audard}, {Bellas-Velidis}, {Benson}, {Berthier},
  {Blomme}, {Burgess}, {Busonero}, {Busso}, {C{\'a}novas}, {Carry}, {Cellino},
  {Cheek}, {Clementini}, {Damerdji}, {Davidson}, {de Teodoro}, {Nu{\~n}ez
  Campos}, {Delchambre}, {Dell'Oro}, {Esquej}, {Fern{\'a}ndez-Hern{\'a}ndez},
  {Fraile}, {Garabato}, {Garc{\'\i}a-Lario}, {Gosset}, {Haigron}, {Halbwachs},
  {Hambly}, {Harrison}, {Hern{\'a}ndez}, {Hestroffer}, {Hodgkin}, {Holl},
  {Jan{\ss}en}, {Jevardat de Fombelle}, {Jordan}, {Krone-Martins}, {Lanzafame},
  {L{\"o}ffler}, {Marchal}, {Marrese}, {Moitinho}, {Muinonen}, {Osborne},
  {Pancino}, {Pauwels}, {Recio-Blanco}, {Reyl{\'e}}, {Riello}, {Rimoldini},
  {Roegiers}, {Rybizki}, {Sarro}, {Siopis}, {Smith}, {Sozzetti}, {Utrilla},
  {van Leeuwen}, {Abbas}, {{\'A}brah{\'a}m}, {Abreu Aramburu}, {Aerts},
  {Aguado}, {Ajaj}, {Aldea-Montero}, {Altavilla}, {{\'A}lvarez}, {Alves},
  {Anders}, {Anderson}, {Anglada Varela}, {Antoja}, {Baines}, {Baker},
  {Balaguer-N{\'u}{\~n}ez}, {Balbinot}, {Balog}, {Barache}, {Barbato},
  {Barros}, {Barstow}, {Bartolom{\'e}}, {Bassilana}, {Bauchet}, {Becciani},
  {Bellazzini}, {Berihuete}, {Bernet}, {Bertone}, {Bianchi}, {Binnenfeld},
  {Blanco-Cuaresma}, {Blazere}, {Boch}, {Bombrun}, {Bossini}, {Bouquillon},
  {Bragaglia}, {Bramante}, {Breedt}, {Bressan}, {Brouillet}, {Brugaletta},
  {Bucciarelli}, {Burlacu}, {Butkevich}, {Buzzi}, {Caffau}, {Cancelliere},
  {Cantat-Gaudin}, {Carballo}, {Carlucci}, {Carnerero}, {Carrasco},
  {Casamiquela}, {Castellani}, {Castro-Ginard}, {Chaoul}, {Charlot}, {Chemin},
  {Chiaramida}, {Chiavassa}, {Chornay}, {Comoretto}, {Contursi}, {Cooper},
  {Cornez}, {Cowell}, {Crifo}, {Cropper}, {Crosta}, {Crowley}, {Dafonte},
  {Dapergolas}, {David}, {David}, {de Laverny}, {De Luise}, \& {De
  March}}]{2023A&A...674A...1G}
{Gaia Collaboration}, {Vallenari}, A., {Brown}, A.~G.~A., {et~al.} 2023, \aap,
  674, A1

\bibitem[{{Gronoff} {et~al.}(2020){Gronoff}, {Maggiolo}, {Cessateur}, {Moore},
  {Airapetian}, {De Keyser}, {Dhooghe}, {Gibbons}, {Gunell}, {Mertens},
  {Rubin}, \& {Hosseini}}]{2020ApJ...890...89G}
{Gronoff}, G., {Maggiolo}, R., {Cessateur}, G., {et~al.} 2020, \apj, 890, 89

\bibitem[{{Guilbert-Lepoutre}(2012)}]{2012AJ....144...97G}
{Guilbert-Lepoutre}, A. 2012, \aj, 144, 97

\bibitem[{{Guilbert-Lepoutre} {et~al.}(2024){Guilbert-Lepoutre}, {Davidsson},
  {Scheeres}, \& {Ciarletti}}]{2024come.book..249G}
{Guilbert-Lepoutre}, A., {Davidsson}, B. J.~R., {Scheeres}, D.~J., \&
  {Ciarletti}, V. 2024, in Comets III, ed. K.~J. {Meech}, M.~R. {Combi},
  D.~{Bockel{\'e}e-Morvan}, S.~N. {Raymodn}, \& M.~E. {Zolensky}, 249--288

\bibitem[{{Gwyn} {et~al.}(2012){Gwyn}, {Hill}, \&
  {Kavelaars}}]{2012PASP..124..579G}
{Gwyn}, S. D.~J., {Hill}, N., \& {Kavelaars}, J.~J. 2012, \pasp, 124, 579

\bibitem[{{Hainaut} \& {Delsanti}(2002)}]{2002A&A...389..641H}
{Hainaut}, O.~R. \& {Delsanti}, A.~C. 2002, \aap, 389, 641

\bibitem[{{Heisler} \& {Tremaine}(1986)}]{1986Icar...65...13H}
{Heisler}, J. \& {Tremaine}, S. 1986, \icarus, 65, 13

\bibitem[{{H{\o}g} {et~al.}(2000){H{\o}g}, {Fabricius}, {Makarov}, {Urban},
  {Corbin}, {Wycoff}, {Bastian}, {Schwekendiek}, \&
  {Wicenec}}]{2000A&A...355L..27H}
{H{\o}g}, E., {Fabricius}, C., {Makarov}, V.~V., {et~al.} 2000, \aap, 355, L27

\bibitem[{{Holt} {et~al.}(2024){Holt}, {Knight}, {Kelley}, {Lister}, {Ye},
  {Snodgrass}, {Opitom}, {Kokotanekova}, {Schwamb}, {Dobson}, {Bannister},
  {Micheli}, {Milam}, {Richardson}, {Gomez}, {Chatelain}, \&
  {Greenstreet}}]{2024PSJ.....5..273H}
{Holt}, C.~E., {Knight}, M.~M., {Kelley}, M. S.~P., {et~al.} 2024, \psj, 5, 273

\bibitem[{{Hui} {et~al.}(2019){Hui}, {Farnocchia}, \&
  {Micheli}}]{2019AJ....157..162H}
{Hui}, M.-T., {Farnocchia}, D., \& {Micheli}, M. 2019, \aj, 157, 162

\bibitem[{{Hui} {et~al.}(2024{\natexlab{a}}){Hui}, {Weryk}, {Micheli}, {Huang},
  \& {Wainscoat}}]{2024AJ....167..140H}
{Hui}, M.-T., {Weryk}, R., {Micheli}, M., {Huang}, Z., \& {Wainscoat}, R.
  2024{\natexlab{a}}, \aj, 167, 140

\bibitem[{{Hui} {et~al.}(2024{\natexlab{b}}){Hui}, {Wiegert}, {Weryk},
  {Micheli}, {Tholen}, {Deen}, {Walker}, \& {Wainscoat}}]{2024ApJ...975L...3H}
{Hui}, M.-T., {Wiegert}, P.~A., {Weryk}, R., {et~al.} 2024{\natexlab{b}},
  \apjl, 975, L3

\bibitem[{{Hung} {et~al.}(2023){Hung}, {Tholen}, {Farnocchia}, \&
  {Spoto}}]{2023PSJ.....4..215H}
{Hung}, D., {Tholen}, D.~J., {Farnocchia}, D., \& {Spoto}, F. 2023, \psj, 4,
  215

\bibitem[{{Ishiguro} {et~al.}(2014){Ishiguro}, {Jewitt}, {Hanayama}, {Usui},
  {Sekiguchi}, {Yanagisawa}, {Kuroda}, {Yoshida}, {Ohta}, {Kawai}, {Miyaji},
  {Fukushima}, \& {Watanabe}}]{2014ApJ...787...55I}
{Ishiguro}, M., {Jewitt}, D., {Hanayama}, H., {et~al.} 2014, \apj, 787, 55

\bibitem[{{Jewitt}(2015)}]{2015AJ....150..201J}
{Jewitt}, D. 2015, \aj, 150, 201

\bibitem[{{Jewitt} {et~al.}(2019){Jewitt}, {Agarwal}, {Hui}, {Li}, {Mutchler},
  \& {Weaver}}]{2019AJ....157...65J}
{Jewitt}, D., {Agarwal}, J., {Hui}, M.-T., {et~al.} 2019, \aj, 157, 65

\bibitem[{{Jewitt} {et~al.}(2021){Jewitt}, {Kim}, {Mutchler}, {Agarwal}, {Li},
  \& {Weaver}}]{2021AJ....161..188J}
{Jewitt}, D., {Kim}, Y., {Mutchler}, M., {et~al.} 2021, \aj, 161, 188

\bibitem[{{Jewitt} {et~al.}(2020){Jewitt}, {Kim}, {Mutchler}, {Weaver},
  {Agarwal}, \& {Hui}}]{2020ApJ...896L..39J}
{Jewitt}, D., {Kim}, Y., {Mutchler}, M., {et~al.} 2020, \apjl, 896, L39

\bibitem[{{Jewitt} \& {Meech}(1986)}]{1986ApJ...310..937J}
{Jewitt}, D. \& {Meech}, K.~J. 1986, \apj, 310, 937

\bibitem[{{Jordi} {et~al.}(2006){Jordi}, {Grebel}, \&
  {Ammon}}]{2006A&A...460..339J}
{Jordi}, K., {Grebel}, E.~K., \& {Ammon}, K. 2006, \aap, 460, 339

\bibitem[{{Joye} \& {Mandel}(2003)}]{2003ASPC..295..489J}
{Joye}, W.~A. \& {Mandel}, E. 2003, in Astronomical Society of the Pacific
  Conference Series, Vol. 295, Astronomical Data Analysis Software and Systems
  XII, ed. H.~E. {Payne}, R.~I. {Jedrzejewski}, \& R.~N. {Hook}, 489

\bibitem[{{Kelley} {et~al.}(2022){Kelley}, {Kokotanekova}, {Holt}, {Protopapa},
  {Bodewits}, {Knight}, {Lister}, {Usher}, {Chatelain}, {Gomez}, {Greenstreet},
  {Angel}, \& {Wooding}}]{2022ApJ...933L..44K}
{Kelley}, M. S.~P., {Kokotanekova}, R., {Holt}, C.~E., {et~al.} 2022, \apjl,
  933, L44

\bibitem[{{Kr{\'o}likowska} \& {Dybczy{\'n}ski}(2010)}]{2010MNRAS.404.1886K}
{Kr{\'o}likowska}, M. \& {Dybczy{\'n}ski}, P.~A. 2010, \mnras, 404, 1886

\bibitem[{{Kulyk} {et~al.}(2018){Kulyk}, {Rousselot}, {Korsun}, {Afanasiev},
  {Sergeev}, \& {Velichko}}]{2018A&A...611A..32K}
{Kulyk}, I., {Rousselot}, P., {Korsun}, P.~P., {et~al.} 2018, \aap, 611, A32

\bibitem[{{Lacerda} {et~al.}(2025){Lacerda}, {Guilbert-Lepoutre},
  {Kokotanekova}, {Inno}, {Epifani}, \& {Snodgrass}}]{2025A&A...697A.210L}
{Lacerda}, P., {Guilbert-Lepoutre}, A., {Kokotanekova}, R., {et~al.} 2025,
  \aap, 697, A210

\bibitem[{{Lang} {et~al.}(2010){Lang}, {Hogg}, {Mierle}, {Blanton}, \&
  {Roweis}}]{2010AJ....139.1782L}
{Lang}, D., {Hogg}, D.~W., {Mierle}, K., {Blanton}, M., \& {Roweis}, S. 2010,
  \aj, 139, 1782

\bibitem[{{Maggiolo} {et~al.}(2020){Maggiolo}, {Gronoff}, {Cessateur}, {Moore},
  {Airapetian}, {De Keyser}, {Dhooghe}, {Gibbons}, {Gunell}, {Mertens},
  {Rubin}, \& {Hosseini}}]{2020ApJ...901..136M}
{Maggiolo}, R., {Gronoff}, G., {Cessateur}, G., {et~al.} 2020, \apj, 901, 136

\bibitem[{{Markwardt}(2009)}]{2009ASPC..411..251M}
{Markwardt}, C.~B. 2009, in Astronomical Society of the Pacific Conference
  Series, Vol. 411, Astronomical Data Analysis Software and Systems XVIII, ed.
  D.~A. {Bohlender}, D.~{Durand}, \& P.~{Dowler}, 251

\bibitem[{{Markwardt} {et~al.}(2023){Markwardt}, {Wen Lin}, {Gerdes}, \&
  {Adams}}]{2023PSJ.....4..135M}
{Markwardt}, L., {Wen Lin}, H., {Gerdes}, D., \& {Adams}, F.~C. 2023, \psj, 4,
  135

\bibitem[{{Meech} \& {Jewitt}(1987)}]{1987A&A...187..585M}
{Meech}, K.~J. \& {Jewitt}, D.~C. 1987, \aap, 187, 585

\bibitem[{{Moreno}(2022)}]{2022Univ....8..366M}
{Moreno}, F. 2022, Universe, 8, 366

\bibitem[{{Nesvorn{\'y}} {et~al.}(2017){Nesvorn{\'y}}, {Vokrouhlick{\'y}},
  {Dones}, {Levison}, {Kaib}, \& {Morbidelli}}]{2017ApJ...845...27N}
{Nesvorn{\'y}}, D., {Vokrouhlick{\'y}}, D., {Dones}, L., {et~al.} 2017, \apj,
  845, 27

\bibitem[{{Park} {et~al.}(2021){Park}, {Folkner}, {Williams}, \&
  {Boggs}}]{2021AJ....161..105P}
{Park}, R.~S., {Folkner}, W.~M., {Williams}, J.~G., \& {Boggs}, D.~H. 2021,
  \aj, 161, 105

\bibitem[{{Raab}(2012)}]{2012ascl.soft03012R}
{Raab}, H. 2012, {Astrometrica: Astrometric data reduction of CCD images},
  Astrophysics Source Code Library, record ascl:1203.012

\bibitem[{{Solontoi} {et~al.}(2012){Solontoi}, {Ivezi{\'c}}, {Juri{\'c}},
  {Becker}, {Jones}, {West}, {Kent}, {Lupton}, {Claire}, {Knapp}, {Quinn},
  {Gunn}, \& {Schneider}}]{2012Icar..218..571S}
{Solontoi}, M., {Ivezi{\'c}}, {\v{Z}}., {Juri{\'c}}, M., {et~al.} 2012,
  \icarus, 218, 571

\bibitem[{{Szab{\'o}} {et~al.}(2007){Szab{\'o}}, {Ivezi{\'c}}, {Juri{\'c}}, \&
  {Lupton}}]{2007MNRAS.377.1393S}
{Szab{\'o}}, G.~M., {Ivezi{\'c}}, {\v{Z}}., {Juri{\'c}}, M., \& {Lupton}, R.
  2007, \mnras, 377, 1393

\bibitem[{{Takada}(2010)}]{2010AIPC.1279..120T}
{Takada}, M. 2010, in American Institute of Physics Conference Series, Vol.
  1279, Deciphering the Ancient Universe with Gamma-ray Bursts, ed. N.~{Kawai}
  \& S.~{Nagataki} (AIP), 120--127

\bibitem[{{Tonry} {et~al.}(2018){Tonry}, {Denneau}, {Flewelling}, {Heinze},
  {Onken}, {Smartt}, {Stalder}, {Weiland}, \& {Wolf}}]{2018ApJ...867..105T}
{Tonry}, J.~L., {Denneau}, L., {Flewelling}, H., {et~al.} 2018, \apj, 867, 105

\bibitem[{{Vere{\v{s}}} {et~al.}(2017){Vere{\v{s}}}, {Farnocchia}, {Chesley},
  \& {Chamberlin}}]{2017Icar..296..139V}
{Vere{\v{s}}}, P., {Farnocchia}, D., {Chesley}, S.~R., \& {Chamberlin}, A.~B.
  2017, \icarus, 296, 139

\bibitem[{{Vokrouhlick{\'y}} {et~al.}(2019){Vokrouhlick{\'y}}, {Nesvorn{\'y}},
  \& {Dones}}]{2019AJ....157..181V}
{Vokrouhlick{\'y}}, D., {Nesvorn{\'y}}, D., \& {Dones}, L. 2019, \aj, 157, 181

\bibitem[{{Waters} {et~al.}(2020){Waters}, {Magnier}, {Price}, {Chambers},
  {Burgett}, {Draper}, {Flewelling}, {Hodapp}, {Huber}, {Jedicke}, {Kaiser},
  {Kudritzki}, {Lupton}, {Metcalfe}, {Rest}, {Sweeney}, {Tonry}, {Wainscoat},
  \& {Wood-Vasey}}]{2020ApJS..251....4W}
{Waters}, C.~Z., {Magnier}, E.~A., {Price}, P.~A., {et~al.} 2020, \apjs, 251, 4

\bibitem[{{Whipple}(1950)}]{1950ApJ...111..375W}
{Whipple}, F.~L. 1950, \apj, 111, 375

\bibitem[{{Willmer}(2018)}]{2018ApJS..236...47W}
{Willmer}, C. N.~A. 2018, \apjs, 236, 47

\bibitem[{{Woodward} {et~al.}(2025){Woodward}, {Rankin}, {Groeller}, {Hogan},
  {Chen}, {Wierzchos}, {Brucker}, {Zhang}, {Wang}, {Kowalski}, {Deen}, {Fay},
  {Leonard}, {Beuden}, {Carvajal}, {Seaman}, {McMillan}, {Larson}, {Nie},
  {Lejoly}, {Mastaler}, {Fuls}, {Gibbs}, {Fazekas}, {Shelly}, {Zhou},
  {Cuillandre}, {:unkn}, {Grauer}, {Sicoli}, {Ventre}, {Colzani}, {Read},
  {Fichtl}, {Rhemann}, {Prosperi}, {Jaeger}, \&
  {Maikner}}]{2025MPEC....D...83W}
{Woodward}, C.~E., {Rankin}, D., {Groeller}, H., {et~al.} 2025, Minor Planet
  Electronic Circulars, 2025-D83

\bibitem[{{Yeomans} {et~al.}(2004){Yeomans}, {Chodas}, {Sitarski}, {Szutowicz},
  \& {Kr{\'o}likowska}}]{2004come.book..137Y}
{Yeomans}, D.~K., {Chodas}, P.~W., {Sitarski}, G., {Szutowicz}, S., \&
  {Kr{\'o}likowska}, M. 2004, in Comets II, ed. M.~C. {Festou}, H.~U. {Keller},
  \& H.~A. {Weaver}, 137

\end{thebibliography}
\bibliographystyle{aa}



\end{document}